\documentstyle[12pt]{article} 

\def\be{\begin{equation}}
\def\ee{\end{equation}}
\def\ben{$$}
\def\een{$$}
\def\ba{\begin{array}{c}}
\def\ea{\end{array}}

\begin{document}

\titlepage
\vspace*{2cm}

\begin{center}{\Large \bf
Perturbation method \\ with triangular propagators \\
 and
 anharmonicities \\ of intermediate strength

 }\end{center}

\vspace{10mm}

\begin{center}
Miloslav Znojil
\vspace{3mm}

\'{U}stav jadern\'e fyziky AV \v{C}R, 250 68 \v{R}e\v{z}, Czech
Republic\\

e-mail: znojil@ujf.cas.cz

\today

\end{center}

\vspace{5mm}

\section*{Abstract}
We propose a new, very flexible version of the
Rayleigh-Schr\"{o}dinger perturbation method which admits a lower
triangular matrix in place of the usual diagonal unperturbed
propagator. The technique and its enhanced efficiency are
illustrated on rational anharmonicities
$V^{(1)}(x)=\beta\times{\rm polynomial}(x)/{\rm polynomial}(x)$.
They are shown tractable, in the intermediate coupling regime, as
${\cal O}(\beta-\beta^{(0)})$ perturbations of exact states at
non-vanishing $\beta^{(0)}\neq 0$. In this sense our method
bridges the gap between the current weak- and strong-coupling
expansions.

\vspace{10mm}

PACS
\hspace{5mm}
03.65.Fd
\hspace{5mm}
03.65.Ge
\hspace{5mm}
 02.60.+y
\hspace{5mm}
 02.70.+d

\newpage

\section{Introduction}

Schr\"{o}dinger equations with anharmonic potentials $ \omega^2x^2
+\beta\,V^{(1)}(x)$ are often solved perturbatively. It is well
known that many practical implementations of this approach are
full of contradictions, well illustrated by the popular quartic
example with $V^{(1)}(x)=x^4$. Its weak-coupling energy estimates
$E (\beta) \approx E (0) + \beta\, E_{(1)} + \ldots + \beta^N
E_{(N)}$ are easily generated via recursion relations \cite{BWu}
but this approximation diverges in the limit $N\to \infty$ at {\em
any} nonzero coupling constant $\beta$ \cite{Dyson}. An
alternative, strong-coupling series in powers of $\beta^{-2/3}$
exists and converges for the sufficiently large $|\beta|$
\cite{Simon}. Unfortunately, the explicit evaluation of its
coefficients is by far not easy \cite{Turbiner}. In the literature
many people have advocated, therefore, a replacement of the
traditional quartic model by a {\em non-polynomial} anharmonicity
\be
V^{(1)}(x) =  {x^2  \over 1 + B\,x^{2}} \equiv
 {1 \over B}\left ( 1 -{1  \over 1 + B\,x^{2}}\right )
  , \ \ \ \ \ \ B >
0. \label{jeho} \label{fless}
 \ee
Its merits belong to the two separate categories. Firstly, its
bounded character enables us to avoid the divergence of the
weak-coupling series. This has been emphasized by several authors
\cite{shortrev}. Secondly, the existence of a few elementary
solutions at certain exceptional couplings $\beta = \beta^{(0)}$
\cite{flessi} enables us to contemplate their perturbations, say,
 \be
 E(\beta) = E (\beta^{(0)}) + \lambda E^{(1)} + \lambda^2 E^{(2)} +
\ldots, \ \ \ \ \ \ \lambda = \beta-\beta^{(0)} \  \label{1}
 \label{expan}
 \ee
in anharmonic regime, near any solvable $\beta^{(0)}\neq 0$
\cite{jasi}. An exceptional character of the latter
intermediate-coupling expansion was its shortcoming. Its
feasibility relied on a replacement of the traditional unperturbed
spectrum by certain auxiliary continued fractions and did not seem
amenable to any sufficiently efficient generalization
\cite{Dubna}.

In the present paper we shall consider the whole class
of the Pad\'{e}-like forces
 \be
V(\beta,x) = \omega^2x^2 + {\beta } \,\left ( \sum_{d=0}^{t}
B_d\,x^{2d}\right )^{-1}\, \sum_{n=0}^{t-1}
 A_n\,x^{2n},
  \ \ \ \ \ \ B_{t} \neq 0. \label{nase}
 \ee
Their {\em strongly} anharmonic perturbation solutions of the type
(\ref{expan}) will be described in full detail. We shall
demonstrate that near many non-vanishing ``intermediate"
couplings $\beta^{(0)}$ our {perturbative approximants} may be
both feasible {\em and} quickly convergent.

The project opens several technical questions.  Firstly, at any
$t \geq 1$, the zero-order solutions must be constructed in
anharmonic regime.  A representative sample of these reference
systems is described thoroughly in Section 2.  It underlines the
real phenomenological appeal of eq. (\ref{nase}) in comparison
with the more popular polynomial models.

At a particular anharmonic $\beta^{(0)} \neq 0$ the usual
construction of a complete unperturbed basis is prohibitively
complicated. After a return to harmonic basis, numerical
integration is needed for evaluation of the necessary matrix
elements of $V^{(1)}(x)$ and the computation of corrections is
difficult even in the lowest order Rayleigh-Schr\"{o}dinger
approximation \cite{Bes}.  Section 3 offers the remedy.
Schr\"{o}dinger equation is represented in a non-orthogonal basis.
Its resulting $(2t+1)-$diagonal matrix form is then much better
accessible to a purely numerical matrix-inversion perturbative
treatment.

In our main Section 4 we deny the latter numerical ``brute force"
philosophy and intend to soften it significantly.  Re-installing
the more traditional recurrent interpretation of perturbation
algorithms we describe a new approach to the Schr\"{o}dinger-like
families of equations with a banded-matrix form of their
Hamiltonians. Our main idea is amply illustrated by its
application to anharmonicities (\ref{jeho}). Its core is a maximal
simplification of the unperturbed propagator ${\cal R}$. In
contrast to its general-matrix form in older methods \cite{jasib}
we shall be able to reduce it to the mere ``half-filled",
triangular matrix.

Section 5 is the summary showing how our new approach opens a way
towards a broader variability of shapes of the theoretical and/or
phenomenological interaction models. With due attention paid to
the non-hermiticity of our (quasi-)Hamiltonian matrices (cf. also
Appendix), our new version of perturbation recipe seems well
prepared for its extensions as well as further practical
computational applications.

\section{Solvable oscillators with $\beta^{(0)} \neq 0$}

\subsection{The simplest model with $t=1$ }

Non-polynomial eq. (\ref{nase}) with $t=1$ is often recalled as
one of the simplest unsolvable anharmonic models in one dimension
\cite{Mitra}. In place of using the differential form of its
Schr\"{o}dinger equation, wave functions $\psi(\beta,x)\in
L_2(-\infty,\infty)$ are expanded in the harmonic (i.e., Hermite
or Laguerre) polynomial basis $\{|n\rangle\}_{n= 0}^\infty$. The
ansatz $\psi(\beta,x)= (1+B\,x^2)\, \sum_{n=0}^\infty\,\langle
x|n\rangle \,h_n $ and scaling $\omega \to 1$ then give
\cite{Whitehead} the three-term recurrences
 \be
\left(
\begin{array}{cccc}
a_0&d_0 &&\\ c_1&a_1&d_1 &\\ &c_2&a_2&\ddots
\\ &&\ddots&\ddots
\ea \right) \, \left( \ba
  h_0\\ h_1\\ h_2\\ \vdots
\ea \right) = 0\  \label{6c}
 \ee
 \ben \ba a_n=\beta+ (1+{B}\,\alpha_n) \, (\varepsilon_n-E)\, , \
\ \ \ \ \ \alpha_n=\varepsilon_n/2 =\langle n |r^2| n
\rangle=2n+\ell+3/2,
\\
d_n =  {B}\,\beta_{n}\,(\varepsilon_n-E) , \ \ \ \ \  \ \beta_n=
\langle n |r^2| n +1\rangle = [(n+1) ( n+\ell+3/2)]^{1/2}
\\
c_n=  {B}\,\beta_{n-1}\,(\varepsilon_n-E) , \ \ \ \ \ \ \ \ \ n =
0, 1, \ldots\ . \ea
 \een
Marginally, let us note that the parity $(-1)^{\ell+1}$ of the
wave functions with $\ell=-1,\ 0$ admits an immediate
re-interpretation as angular momentum $\ell= 0, 1,\ldots$ in three
dimensions with the regularity $ \psi(\beta,r) \sim r^{\ell+1}$ of
the radial wave functions near the origin \cite{comm}.

\subsubsection{Termination conditions and exact solutions}

The existence of the terminating exact solutions of eq. (\ref{6c})
is well known \cite{flessi}. Let us mark them by a superscript
$^{(0)}$. With normalization $h_q^{(0)}\neq 0$ their termination
property
 \be
 h_{q+1}^{(0)}= h_{q+2}^{(0)} =\ldots =0
  \label{terminuj}
 \ee
requires that $c_{q+1}^{(0)}=0$. This means that $E^{(0)} =
4q+2\ell+7 \equiv \varepsilon_{q+1}$. The related energy is not
arbitrary. Vice versa, bound states $\psi(\beta,x)$ with $E \neq
E^{(0)}$ have to be defined by infinite series \cite{cf}. In one
dimension a sample of their spectrum is given in Table 1. It
indicates that with a growth of the barrier the low-lying energies
merge in almost degenerate doublets with opposite parities.
According to Figure~1 a very good fit of these numerical values
$E=E(\beta)$ is provided by parabolas. One may expect that besides
our ansatz (\ref{expan}) a useful methodical alternative could be
also sought in perturbative expansions of couplings
$\beta=\beta(E)$ and of the related Sturmian wave functions
\cite{Whitehead,Sturmians}.

Even for the terminating bound states with $E = E^{(0)}$ we have
to guarantee that the secular determinants vanish. Up to $q=3$ the
latter condition is non-numerical. For illustration we may fix
$B=B^{(0)}=1$ and choose the even parity $\ell = -1$. Then we get
the elementary implicit polynomial definitions
\be
\ba y-6=0, \ \ \ \ \ \ \ \ \ \ q = 0\\ y^2-26\,y + 152 = 0, \ \ \
\ \ \ \ \ q=1\\ y^3-68\,y^2+1372\,y-8304, \ \ \ \ \ \ \ q=2\\
y^4-140\,y^3+6588\,y^2-123216\,y+777600= 0, \ \ \ \ \ q=3 \ea
\label{flessass}
 \ee
of the partially
solvable couplings $y=y(q) \equiv \beta^{(0)}$.
Besides their $q=0$ (linear), $q=1$ (quadratic) and $q=2$
(Cardano) explicit solutions we may write down all the four
exact $q=3$ roots
 \ben \beta^{(0)}=35 +\varepsilon_1\,
\sqrt{127-2\,\sqrt{6821}\,\cos \left( {1 \over 3}{\Theta} + {\pi
\over 6}\right)} +\varepsilon_2\, \sqrt{127+2\,\sqrt{6821}\,\sin
\left( {1 \over 3} {\Theta} + {\pi \over 3}\right) } \een \ben
 +\varepsilon_3\,
\sqrt{127-2\,\sqrt{6821}\,\sin  {1 \over 3}{\Theta} }, \ \ \ \ \ \
\ {\Theta} = {\rm arctg}{9\,\sqrt{63908723442661575155} \over
45902084710}
 \een
with $ (\varepsilon_1,\varepsilon_2,\varepsilon_3) = (-,-,+),
(-,+,-), (+,-,-)$ and $ (+,+,+)$.

All our termination-compatible $q \leq 3$ values of the coupling
$\beta^{(0)}$ remain real (cf. their list in Table~2). All the
related forces acquire a double well shape since all our roots
satisfy its sufficient condition $\beta^{(0)} > 1$.

\subsection{Unperturbed solutions with $t = 2$}

The $t=2$ option in (\ref{nase}) gives the ``first nontrivial"
potential
 \be
V(x) = x^2 + {\mu \,x^2 + \nu \,\over \left (1-g\,x^2 \right )^2 +
f\,x^2 }\ . \label{imple}
 \ee
{\it Mutatis mutandis}, equation (\ref{6c}) becomes replaced by
the $q+3$ relations
 \be
\left(
\begin{array}{cccccc}
a_0&d_0^{[1]}&d_0^{[2]}&&&\\
 c_1^{[1]}&a_1&d_1^{[1]}&d_1^{[2]}&&\\
c_2^{[2]}& c_2^{[1]}&a_2&d_2^{[1]}&d_2^{[2]}&\\
 &\ddots&\ddots&\ddots&\ddots&\ddots\\
 &&c_{q-1}^{[2]} &c_{q-1}^{[1]} &a_{q-1}&d_{q-1}^{[1]}\\
  &&&c_{q}^{[2]} &c_{q}^{[1]} &a_{q}\\
&&&&c_{q+1}^{[2]}&c_{q+1}^{[1]}\\ &&&&&c_{q+2}^{[2]}
 \ea \right) \, \left(
\ba h_0{^{(0)}}\\ h_1{^{(0)}}\\ \vdots \\ h_q{^{(0)}} \ea \right)
= 0 \ .\label{6cdt2}
 \ee
The last row implies that
$c_{q+2}^{[2]}= 0$ and fixes the energy $E^{(0)} = 4q+2\ell+11$.
The coupled rest remains over-determinate and defines the
$q+1$ unknown coefficients (normalized, say, to $h_q^{(0)}=1$) and
two coupling constants. At $q=0$
we have
\be
\nu^{(0)}=(4\ell+6)(f-2g)+8, \ \ \ \ \ \ \mu^{(0)} =
(8\ell+20)g^2+4f-8g\ \ \ \ \ \ q=0\ . \label{sample}
 \ee
These couplings are real for all the parameters
$f$, $g$ and $\ell$.

For the sake of brevity, let us put $f=g=1$ from now on. With
option $\ell=0$ (in one or three dimensions)
and degree $q=1$ our
conditions (\ref{6cdt2}) degenerate to the cubic
equation $\nu^3+48\,\nu -360=0$. Its only real Cardano root
\be
\nu^{(0)}= \left(4\,\sqrt{2281}+180\right)^{1/3}
-\left(4\,\sqrt{2281}-180\right)^{1/3} \approx 4.959 146 611 331
66 \ \label{amp}
 \ee
with $\mu^{(0)} \approx 14.941 997 536 546$ and
normalization $h_1^{(0)}=1$ leads to the exact
\be
h_0^{(0)}= -\left({\sqrt{13686} \over 108}+{653\,\sqrt{6} \over
486 }\right)^{1/3}- \left({653\,\sqrt{6} \over 486} -{\sqrt{13686}
\over 108}\right)^{1/3}-{2\,\sqrt{6} \over 9}
 \ee
i.e., $h_0^{(0)} \approx -3.481 950 172 214 96$.
The related energy $E^{(0)} = 15$ corresponds to the first excited
state in $s-$wave.  Its ground state predecessor does not
terminate, $q \to \infty$. For it, the Runge-Kutta integration
gives $E_{gs} \approx 10. 943 408 413$.

At the next choice of $q=2$ with the quasi-harmonic energy
$E^{(0)}=19$ and with the same convenient normalization
$h_2^{(0)}=1$ equation
(\ref{6cdt2}) is fully pentadiagonal,
\be
\left(
\begin{array}{ccc}
{3 \over 2}\,\mu+\nu-52 &{\sqrt{6} \over 2}\,\mu-32\,\sqrt{6}
&-8\,\sqrt{30}\\ {\sqrt{6} \over 2}\,\mu-24\,\sqrt{6} &{7 \over
2}\,\mu+\nu-195 &\sqrt{5}\,\mu-96\,\sqrt{5}
\\
-4\,\sqrt{30} &\sqrt{5}\,\mu-64\,\sqrt{5} &{11 \over
2}\,\mu+\nu-330\\0& -2\,\sqrt{210} &{\sqrt{42} \over
2}\,\mu-24\,\sqrt{42} \ea \right) \left( \ba h_0^{(0)}\\
h_1^{(0)}\\ h_2^{(0)} \ea \right) = 0. \label{effek}
 \ee
In an ascending order of its rows we eliminate
 \ben \mu=4\, (\sqrt{5}\,h_1^{(0)}+12),\ \ \
\ \nu=2\,[2\,\sqrt{30}\,h_0^{(0)} -10\,(h_1^{(0)})^2
+3\,(11-\sqrt{5}\,h_1^{(0)})] \een and \ben h_0^{(0)}=\sqrt {30}\,
(20\,(h_1^{(0)})^3 -8\,\sqrt{5}\,(h_1^{(0)})^2 -59\,h_1^{(0)}
+48\,\sqrt {5}) /(180\,h_1^{(0)}).
 \een
With $h_1^{(0)}=y\,\sqrt{5}$ our problem degenerates to the single
sextic polynomial equation with integer coefficients, \ben
2500\,y^6+1000\,y^5-7125\,y^4+100\,y^3+5065\,y^2-264\,y-1152=0.
\een Its two real roots are easily localized numerically, $
y_1\approx -1.69830$ and $ y_2\approx 0.815078 $.  A close analogy
with $t=1$ is preserved.  With the same ease we may generate the
two $t=2$ oscillators from a tenth-degree polynomial at $q=3$ (see
Table~3) etc. The related energies were evaluated numerically.
Their sample is given in Table 4. They safely stabilize at cut-off
$M = 15$.

\subsection{$t =3$ and more}

At any $t > 1$ the requirement (\ref{terminuj}) leads to the
$(t+q+1)\times (q+1)-$dimensional generalization of eq.
(\ref{6cdt2}). Its last, decoupled condition $c_{q+t}^{[t]}= 0$ is
satisfied if and only if
$E^{(0)} = 4t+4q+2\ell+3$.
The remaining $t+q$ coupled equations
\be
\left(
\begin{array}{ccccccc}
a_0&d_0^{[1]}&\ldots&d_0^{[t]}&0&\ldots& 0\\
c_1^{[1]}&a_1&\ldots&d_1^{[t-1]}&d_1^{[t]} &\ldots&0\\
&&\ldots&&&&\\ 0&&\ldots&0&c_{t+q-2}^{[t]} &c_{t+q-2}^{[t-1]}
&c_{t+q-2}^{[t-2]}\\
0&&\ldots&0&0&c_{t+q-1}^{[t]}&c_{t+q-1}^{[t-1]} \ea \right) \,
\left( \ba h_0{^{(0)}}\\ h_1{^{(0)}}\\ \vdots \\ h_q{^{(0)}} \ea
\right) = 0 \label{6cdt}
 \ee
determine all the $q$ normalized projections $h_j^{(0)}$ plus $t$
parameters in potential itself. With the growth of $t$ the
selfconsistent search for these exact solutions becomes less and
less straightforward. Due to the implicit nonlinearity of eq.
(\ref{6cdt}) we must verify that its solutions keep the forces
real and non-singular. Both these properties have to be verified
{\it a posteriori}.

It is useful to notice that at $q=0$ the explicit solutions remain
elementary at any index $t \geq 1$.  At $t=3$ the purely
non-numerical solutions still exist at $q>0$.  This is slightly
unexpected. For illustration, let us employ the
quartic-over-sextic model
\be
V(\beta^{(0)},r) = x^2 + {u^{(0)}+v^{(0)}\,x^2+w^{(0)}\,x^4 \over
1+ x^6}\ . \label{nubip}
 \ee
In a search for its symmetric bound states in one dimension
($\ell=-1$) the ``first nontrivial" choice of $q=1$ gives
$E^{(0)}=17$. The abbreviation $h_0^{(0)}=a$ and the eliminations
guided by our previous experience re-parametrize the couplings, $$
w^{(0)}=2\,(\sqrt{2}\,a+27), \ \ \ \ \
v^{(0)}=-\sqrt{2}\,(\sqrt{2}\,a^2+11\,a-6\,\sqrt{2}) $$ $$
u^{(0)}=\sqrt{2}\,(2\,a^3+10\,\sqrt{2}\,a^2-23\,a+18\,\sqrt{2})/2.
 $$
The whole algebra degenerates to the single equation in
$y=\sqrt{2}\,a$, $$ y^4+9\,y^3-33\,y^2+27\,y-12= 0, \ \ \ \ \
\ell=-1, \ q=1, \ \ t = 3.
 $$
Two of its roots are complex, $ y_{3,4}=0.451\pm 0.534\,i $, and
the real doublet is given by the expression
$$ y_{1,2}= -{9 \over 4}- \sqrt{{169\over 16}
-\sqrt[3]{\sqrt{25057}+161 \over 16}-\sqrt[3]{161- \sqrt{25057}
 \over 16}} \pm Y_a $$ where $$ Y_a=\sqrt{{169\over 8} +\sqrt[3]{
\sqrt{25057}+161 \over 16}+\sqrt[3]{161- \sqrt{25057} \over
16}+\left| Y_b \right|}.
 $$
The last item is a positive square root of another sum, $$
(Y_b)^2={ 28177\over 64} + \sqrt[3]{161\, \sqrt{25057}+25489\over
128}+\sqrt[3]{25489-161\, \sqrt{25057}\over 128} $$ $$
+\sqrt[3]{4347\, \sqrt{25057}+688203\over
128}+\sqrt[3]{688203-4347\, \sqrt{25057} \over 128} $$ $$
+\sqrt[3]{4826809\, \sqrt{25057}+777116249\over
1024}+\sqrt[3]{777116249 -4826809\,\sqrt{25057}\over 1024}.
  $$
In the second illustration with $q=2$, $E^{(0)}=21$ and
abbreviations $ h_0 ^{(0)}=a=y/\sqrt{6}$, $h_1
^{(0)}=b=z/\sqrt{3}$ and $ h_2 ^{(0)}=1 $ the eliminations $$
w^{(0)}=2\,\sqrt{3}\,(2\,b+13\,\sqrt{3}), \ \
\
v^{(0)}=2\,\sqrt{3}\,(2\,\sqrt{2}\,a-2\,\sqrt{3}\,b^2-7\,b+12\,\sqrt{3})
$$ $$ u^{(0)}=-3\,(a\,(6\,\sqrt{2}\,b
+4\,\sqrt{6}-4\,\sqrt{3}\,b^3-4\,b^2+31\,\sqrt{3}\,b-64) $$  $$
y={24\,z^5+4\,z^4-674\,z^3+1437\,z^2+1355\,z-5292 \over
2\,(20\,z^3+47\,z^2-117\, z-604)} $$ generate the tenth-degree
polynomial in $z$ possessing the four real roots. These results
are summarized in Table~5.

\section{Perturbations}

Any family of phenomenological potentials may be supposed
approximated by an asymptotically harmonic Pad\'{e} approximant
(\ref{nase}) of a suitable degree $t$. Up to the $K-$th order a
consistency of its subsequent perturbative treatment is guaranteed
whenever the remainder is kept sufficiently small, $ V^{\rm
(phenomenological)}(x) - V^{\rm (Pad\acute{e})}(x) = {\cal
O}(\lambda_{\rm min}^{K+1}) = {\cal O}(\lambda_{\rm max}^{K+1}) $.
The stability of approximation requires that the more or less
random poles in $ V^{\rm (Pad\acute{e})}(x)$ are under firm
control.  In the case of our class of forces (\ref{nase}) this is
most easily achieved by their unique~\cite{Korn} partial-fraction
re-arrangement \be V(\beta,x) =\omega^2x^{2}+ \sum_{m=1}^{M_1}
\sum_{j=1}^{J(m)} {\sigma_{mj} \over \left( 1+e_m x^2 \right)^j} +
\sum_{n=1}^{N_2} \sum_{k=1}^{K(n)} {\mu_{nk} x^2 +\nu_{nk} \over
\left[ 1 + (f_n-2g_n) x^2+ g_n^2 x^4 \right]^k }\ .
\label{potenciala}
 \ee
An instructive illustration is offered by the $t=2$ example
(\ref{imple}). The maximal admissible range of perturbation of
its couplings must be restricted by the condition of positivity
of the denominator.  This means that we must have $f=f(\lambda)>
0$ or $f(\lambda)=0$ and $g=g(\lambda) \leq 0$ or $0>
f(\lambda)> 4g(\lambda)$ for  $\lambda \in (\lambda_{\rm
min},\lambda_{\rm max})$.

We shall expand the perturbed wave functions in the same modified
oscillator basis as above,
 \be
{\psi(\lambda,r) = {\cal B}(r)} \sum_{n=0}^{\infty}\,h_n(\lambda
)\,\langle r | n \rangle \equiv \sum_{n=0}^{\infty}\,h_n(\lambda
)\,\langle r | \Xi_n \rangle, \ \ \ \ \
 {\cal B}(r) =
\sum_{d=0}^{t} B_d\,x^{2d} . \label{inwavef}
 \ee
This leads to the infinite-dimensional
Schr\"{o}dinger equation
\be
[{\cal H}(\lambda)-E(\lambda){\cal D}(\lambda)]\vec{h}(\lambda)=0.
\label{sperto} \label{6cd}
 \ee
It degenerates back to the finite-dimensional problems of
preceding section in the unperturbed limit $\beta \to
\beta^{(0)}$.

\subsection{Phenomenological appeal of Pad\'e oscillators}

The degree $t$ in eq. (\ref{nase}) specifies also a half-bandwidth
of our $(2t+1)-$diagonal quasi-Hamiltonians $ [{\cal H}(\lambda) -
E(\lambda){\cal D}(\lambda) ]$. This $t=t[J(1), \ldots, J(
M_1),\,K(1), \ldots, K(N_2)]$ grows rather quickly with all its
arguments. Vice versa, the very first $t$'s already offer a rich
variety of possible shapes of the phenomenological potential
(cf. Figures 2 and 3).

In the simplest $t=1$ example (\ref{fless}) the numerically
calculated $\lambda-$ or $\beta-$dependence of energies exhibits a
roughly quadratic shape, $\beta \sim (E+const)^2+const$.  For the
four lowest states this is illustrated by Figures 1 and 4. The
exceptional exact energies (marked by crosses) are scattered all
over the coupling-energy plane. These quasi-harmonic points may be
inter-connected by auxiliary lines (cf. Figures 5 and 6 and ref.
\cite{Gallas}). In a way, these lines generalize the harmonic
spectrum to $\beta^{(0)} \neq 0$. Their distinguished feature
seems to be an asymptotically almost equidistant and almost linear
shape, resembling strongly their harmonic predecessor.

Only the first few energies exhibit in fact a pronounced
non-equidistant spacing.  The onset of the almost equidistant
behaviour moves only slowly up with the growth of $\beta^{(0)}$.
The approximate linearity of dependence of the $n-$th energy level
$E_n$ on the value of the coupling $\beta$ is remarkable. We may
expect that the first-order perturbation formulae will reproduce
the $t=1$ energies $E(\lambda)$ with decent precision in a broad
interval of their $\lambda-$dependence.

It is well known that in the context of studies of double wells
one of the big challenges to perturbation theory is posed by the
related approximate degeneracy between the even and odd states. An
explicit illustration of this phenomenon is provided by Table 1.
It indicates that the long-lasting puzzle of perturbations of the
quasi-degenerate spectra in the deep double wells may find one of
its very natural resolutions in the present language since our
formalism treats the states with different parity as perturbations
of different systems.  For example, in between the first two
couplings of the Table [i.e., for $\beta \in (64.89\ldots,
81.88\ldots)$] one should calculate the ground state energies as
perturbations of $E^{(0)}=17$ while the very close first
excitations should be perturbations of $E^{(0)}=19$ in another
potential.  In low orders the split of energies will probably
remain disguised by errors of their separate perturbative
determinations but the related quasi-degenerate eigenfunctions
themselves become clearly distinguished by their parity.

\subsection{Rayleigh-Schr\"{o}dinger expansions near
$\beta^{(0)}$}

The measure $\lambda=\beta-\beta^{(0)}$ of
deviation of our perturbed Schr\"{o}dinger equation from its
zero-order form should be sufficiently small in the Kato's
sense \cite{Kato}.  Then, any analytic $\lambda-$dependence of the
matrices
 \be
 {\cal D}(\lambda)= {\cal D}(0) + \lambda {\cal D}^{(1)} +
\lambda^2 {\cal D}^{(2)} + \ldots, \ \ \ \ \ \ \ {\cal
H}(\lambda)= {\cal H}(0) + \lambda {\cal H}^{(1)} + \lambda^2
{\cal H}^{(2)} + \ldots \ ,  \label{dalsi}
 \ee
may be expected to imply the validity of the energy series
(\ref{1}) and of its wave function counterpart
 \be
h_j =h_j (\lambda) = h_j{(0)} + \lambda h_j^{(1)} + \lambda^2
h_j^{(2)} + \ldots \ .\label{wavef }
 \ee
This transforms our $\lambda-$dependent Schr\"{o}dinger equation
into a set of its separate ${\cal O}(\lambda^k)$ components. At $k
= 0$, the unperturbed problem $[{\cal H}(0) - E(0){\cal
D}(0)]\vec{h}{(0)} = 0$ of preceding section is re-obtained.  Next
we get its ${\cal O}(\lambda)$ descendant
\be
[{\cal H}(0) - E^{(0)}{\cal D}(0)] \ \vec{h}^{(1)} = [
E^{(0)}{\cal D}^{(1)} -{\cal H}^{(1)} ]\ \vec{h}{(0)} +
E^{(1)}{\cal D}(0) \vec{h}{(0)}\ . \label{first}
 \ee
In compact notation with abbreviations $ \vec{\tau}^{(0)} \equiv
[E^{(0)}{\cal D}^{(1)} -{\cal H}^{(1)} ]\vec{h} {(0)}$,
$\vec{\rho}^{\{0\} } \equiv {\cal D}(0)\vec{h}{(0)}$ and ${\cal M}
= {\cal H}(0) - E(0){\cal D}(0)$ this equation shares its form
with all the subsequent ${\cal O}(\lambda^k)$  equations
 \be
{\cal M}  \vec{h}^{(k)} = \vec{\tau}^{(k-1)} + E^{(k)}
\,\vec{\rho}^{\{0\} }\  \label{oureq}
 \ee
requiring only the further abbreviation
 \ben \vec{\tau}^{(1)}= [ E^{(0)}{\cal
D}^{(2)}+ E^{(1)}{\cal D}^{(1)} -{\cal H}^{(2)} ]\vec{h}{(0)} + [
E^{(0)}{\cal D}^{(1)}+ E^{(1)}{\cal D}(0) -{\cal H}^{(1)}
]\vec{h}^{(1)}
 \een
and, in general, \ben \vec{\tau}^{(k-1)}= \left[  \sum_{j=0}^{k-1}
E^{(j)}{\cal D}^{(k-j)}-{\cal H}^{(k)} \right]\vec{h}{(0)} + \een
\be +\sum_{m=1}^{k-1} \left[ \sum_{i=0}^{k-m-1} E^{(i)}{\cal
D}^{(k-m-i)} + E^{(k-m)}{\cal D}(0)
 -{\cal
H}^{(k-m)} \right]\vec{h}^{(m)}\ . \label{taurek}
 \ee
As long as ${\rm det} {\cal M} = 0$ each particular solution
${\vec{h}}^{(k)} $ may contain an arbitrarily large admixture of
the zero-order column vector $\vec{h}{(0)}$.  This is the well
known renormalization freedom of perturbative wave functions in
quantum mechanics. We get rid of it by the normalization
$h_q^{(k)} = 0$ in each perturbation order $k > 0$. Such a
convention differs from the standard textbook recommendations
but serves the same purpose and makes the solutions of our key
eq. (\ref{oureq}) well defined.

\subsection{Example: tridiagonal ${\cal M}$}

Let us choose $t=1$, fix the nodal count $q$ and accept an exact
solution at $\beta=\beta^{(0)}$ as our illustrative
$^{(0)}-$superscripted zero-order approximation. The perturbed
couplings $\beta =\beta^{(0)}+\lambda$ and
$B=B^{(0)}+\lambda\,B^{(1)} + \ldots$ with a small measure of
perturbation $\lambda \neq 0$ enter the infinite dimensional
Schr\"{o}dinger equation (\ref{6c}) or (\ref{sperto}). After
appropriate insertions we get the $t=1$ set of equations
(\ref{oureq}),
 \be
\left(
\begin{array}{cccc}
a^{(0)}_0&d^{(0)}_0 &&\\ c^{(0)}_1&a^{(0)}_1&d^{(0)}_1 &\\
&c^{(0)}_2&a^{(0)}_2&\ddots
\\ &&\ddots&\ddots
\ea \right) \, \left( \ba
  h^{(k)}_0\\ h^{(k)}_1\\ h^{(k)}_2\\ \vdots
\ea \right) = E^{(k)}\, \left( \ba
  \rho^{(0)}_0\\ \rho^{(0)}_1\\ \rho^{(0)}_2\\ \vdots
\ea \right)
 +
\left( \ba
  \tau^{(k-1)}_0\\ \tau^{(k-1)}_1\\ \tau^{(k-1)}_2\\ \vdots
\ea \right) \ .\label{6cper}
 \ee
Only the first $q+2$ components of $\vec{\rho}^{(0)}={\cal
D}^{(0)}\,\vec{h}^{ (0)}$ are nonzero since
\be
\left( \ba
  \rho^{(0)}_0\\ \rho^{(0)}_1\\ \rho^{(0)}_2\\ \vdots
\ea \right)= \left(
\begin{array}{cccc}
a^{\{kin\}}_0&d^{\{kin\}}_0 &&\\
c^{\{kin\}}_1&a^{\{kin\}}_1&d^{\{kin\}}_1 &\\
&c^{\{kin\}}_2&a^{\{kin\}}_2&\ddots
\\ &&\ddots&\ddots
\ea \right) \, \left( \ba
  h^{(0)}_0\\ h^{(0)}_1\\ h^{(0)}_2\\ \vdots
\ea \right) , \label{tenvec} \ee \ben a^{\{kin\}}_n=
(1+{f^{(0)}}\,\alpha_n) , \ \ \ \ \ \ \ \ d^{\{kin\}}_n =
{f^{(0)}}\,\beta_{n}, \ \ \ \ \ \ \ \ \ c^{\{kin\}}_n=
{f^{(0)}}\,\beta_{n-1}\ .
 \een
In contrast, the compressed previous-order (i.e., already known)
corrections $\tau_j^{(k-1)}$ only terminate in the first order, at
$k=1$.

\subsubsection{Ground-state illustration}

In the $k=1$ and $q=0$ exemplification of equation (\ref{6cper})
 \be
\left(
\begin{array}{cc|ccc}
0&d_0^{(0)}&0&\ldots&\\ 0&a_1^{(0)}&0&0&\ldots\\ \hline
 0&
c_2^{(0)}&a_2^{(0)}&d_2^{(0)}&\ldots\\ 0&
0&c_{3}^{(0)}&a_{3}^{(0)}&\ldots\\ \vdots&&&\ddots&\ddots \ea
\right) \, \left( \ba 0\\ h_1^{(1)}\\ \hline h_2^{(1)}\\
h_3^{(1)}\\ \vdots \ea \right)
 =
\left( \ba
  \tau_0^{(0)}\\ 0\\
\hline 0\\ 0\\ \vdots \ea \right)
 +E^{(1)} \
\left( \ba
  \rho_0^{\{0\}}\\ \rho_1^{\{0\}}\\
\hline 0\\ 0\\ \vdots \ea \right) \  \label{firdd}
 \ee
the first two energy-dependent rows decouple from the rest. For
the even-parity $\ell = -1$  they read
 \ben \ba
-2\,\sqrt{2}\,h_1^{(1)}=-\beta^{(1)}+3\,E^{(1)}/2,\\
\beta^{(0)}\,h_1^{(1)}=E^{(1)}/\sqrt{2}\ . \ea
 \een
As long as $\beta^{(0)} = 6$ and $\beta^{(1)} = 1$ their solution
reproduces the current textbook first-order overlap formula for
the energy,
 \be
E^{(1)} = { \int_{-\infty}^\infty \exp (-x^2) (1+x^2) dx \over
\int_{-\infty}^\infty \exp (-x^2) (1+x^2)^2 dx } = {6 \over 11} \
.\label{RSian}
 \ee
This test demonstrates the user-friendliness of our
non-Hermitian recipe.

\subsubsection{The first excitation}

In our preceding illustration we did not mark the cut-off $M$ in
${\cal M}$. For the next, $q=1$ state with $M=6$ (chosen small for
paedagogical purposes) we have to solve the set
 \be
\left(
\begin{array}{ccc|cccc}
a_0{}&0{}&0&0&0&0&0\\ c_1{}&0{}&d_1{}&0&0 &0&0\\ 0&0&a_2{}&0&0
&0&0\\ \hline 0& 0&c_{3}{}&a_{3}{}&d_{3}{}&0&0\\ 0&0&
0&c_{4}{}&a_{4}{}&d_{4}{}&0
\\ 0&0&0&0&c_{5}{}&a_{5}{}&d_{5}{}
\\0& 0&0&0&0&c_{6}{}&a_{6}{}
 \ea \right)
\, \left( \ba  h_0^{(k)}\\0\\ h_2^{(k)}\\ \hline
 h_3^{(k)}\\ h_4^{(k)}\\ h_5^{(k)} \\ h_6^{(k)}
\ea \right)
 =
\left( \ba
  \tau_0^{(k-1)}\\ \tau_1^{(k-1)}\\ \tau_2^{(k-1)}\\
\hline \tau_3^{(k-1)}\\ \tau_4^{(k-1)}\\ \tau_5^{(k-1)}\\
\tau_6^{(k-1)} \ea \right)
 +E^{(k)} \
\left( \ba
  \rho_0^{\{0\}}
\\ \rho_1^{\{0\}}\\
 \rho_2^{\{0\}}\\
\hline 0\\ 0\\ 0 \\ 0 \ea \right) \ . \label{llcdd}
 \ee
The upper part of this equation (separated by the inset lines)
stays decoupled.  We can underline that in contrast to the current
textbook recipe we do not need any left eigenvector of ${\cal M}$
and still can define the $k-$th energy correction via the finite,
$(q+t+1)-$dimensional matrix inversion
\be
\left(
\begin{array}{ccc}
 a_0{}& \rho_0^{\{0\}} &0\\ c_1{}&\rho_1^{\{0\}} &d_1{}\\
0&\rho_2^{\{0\}} &a_2{} \ea \right) \, \left( \ba h_0^{(k)}\\
-E^{(k)}
\\
h_2^{(k)} \ea \right)
 =
\left( \ba
  \tau_0^{(k-1)}\\ \tau_1^{(k-1)}\\ \tau_2^{(k-1)}
\ea \right)\ . \label{kkdd}
 \ee
In the second, lower part of eq.  (\ref{llcdd}) the cut-off $M$
should grow to infinity in principle.  The matrix-inversion
evaluation of the remaining wave function components
$h_{q+t+1}^{(k)}, \ h_{q+t+2}^{(k)},\ldots$ is much more difficult
and acquires a purely numerical character. In a schematic
representation $\vec{h}^{(k)} \sim {\cal R} \vec{\tau}^{(k-1)} +
\ldots$ the  ``unperturbed propagator" $ {\cal R}$ is a general,
{\em fully non-diagonal} matrix.

\subsubsection{The higher excitations}

The size of the upper part of eq. (\ref{6cper}) grows with $q$
(cf. (\ref{firdd}) and (\ref{llcdd})). Only its last,
$(q+1)-$subscripted line remains trivial. As long as
$c_{q+1}^{(0)}=0$, $d_{q+1}^{(0)}=0$ and
$a_{q+1}^{(0)}=g^{(0)}>0$, it degenerates to an explicit
definition of the $(q+1)-$st coefficient in terms of the not yet
specified energy,
\be
h_{q+1}^{(k)} ={E^{(k)}\,\rho_{q+1}^{(0)}+\tau_{q+1}^{(k-1)} \over
g^{(0)} } \label{fixing}.
 \ee
Thus, in practice, our equation (\ref{6cper}) decays in the two
equally difficult subsystems at the larger $q$.  Its upper $q+1$
rows are, fortunately, not as complicated as they look. Firstly,
an immediate elimination of energies may help for $q \gg 1$. It
proves unexpectedly easy since the array $\vec{\rho}^{(0)}$ is a
left eigenvector of our non-Hermitian quasi-Hamiltonian ${\cal
M}$. The left action of this vector on eq.  (\ref{oureq})
eliminates all $h_n^{(k)}$'s and defines the $k-$th energy
correction at any $t$,
\be
E^{(k)} = -{\sum_{m=0}^{q+t}
 \rho^{(0)}_m
\tau^{(k-1)}_m \over \sum_{n=0}^{q+t} \left( \rho^{(0)}_n
\right)^2 }.
 \label{knowne}
 \ee
With this knowledge, equation (\ref{fixing}) determines
$h_{q+1}^{(k)}$. At the exact energy (\ref{knowne}) an abbreviated
right-had-side vector $\tilde{\tau}^{(k-1)}_m \equiv
{\tau}^{(k-1)}_m+ E^{(k)}\,\rho^{(0)}_m$, $m = 0, 1,\ldots$ enters
the equation  (\ref{6cper}) whose first $q+1$ separate rows become
linearly dependent.  We omit the very first one as redundant.
Simultaneously, our normalization $h^{(k)}_{q}=0$ annihilates a
column in ${\cal M}$ and we get the reduced equation
\be
\left(
\begin{array}{ccccc}
c^{(0)}_1& a^{(0)}_1&d^{(0)}_1 &&\\ &c^{(0)}_2&
a^{(0)}_2&d^{(0)}_2 &\\ &&\ddots&\ddots &\ddots\\ &
&&c^{(0)}_{q-1}& a^{(0)}_{q-1}\\ & &&&c^{(0)}_{q} \ea \right) \,
\left( \ba
  h^{(k)}_0\\ h^{(k)}_1\\
\vdots\\ h^{(k)}_{q-2}\\ h^{(k)}_{q-1} \ea \right) = \left( \ba
 \tilde{ \tau}^{(k-1)}_1\\  \tilde{ \tau}^{(k-1)}_2\\ \vdots\\
\tilde{ \tau}^{(k-1)}_{q-1}\\ \tilde{ \tau}^{(k-1)}_{q} \ea
\right)  \label{7rfin}.
 \ee
This gives a non-diagonal, upper triangular generalization
\be
\left( \ba
  h^{(k)}_0\\ h^{(k)}_1\\
\vdots\\ h^{(k)}_{q-2}\\ h^{(k)}_{q-1} \ea \right) =
\left(
\begin{array}{ccccc}
1/c^{(0)}_1& {\cal R}_{12}&\ldots&& {\cal R}_{1q}\\
&1/c^{(0)}_2& {\cal R}_{23}&\ldots& {\cal R}_{2q}\\
&&\ddots&\ddots &\vdots\\
&&&1/c^{(0)}_{q-1}& {\cal R}_{q-1q}\\
&&&&1/c^{(0)}_{q}
 \ea \right) \,  \left( \ba
 \tilde{ \tau}^{(k-1)}_1\\  \tilde{ \tau}^{(k-1)}_2\\ \vdots\\
\tilde{ \tau}^{(k-1)}_{q-1}\\ \tilde{ \tau}^{(k-1)}_{q} \ea
\right)  \label{8fin}
 \ee
of the current rule based on a diagonal unperturbed propagator.
Unfortunately, any extension of the latter trick fails. For all
$t> 1$ a different approach is needed.

\section{New recipe using triangular propagators}

Previous examples clarified an exceptional role of the $(q+t)-$th
row in ${\cal M}$ where $c_{q+t}=d_{q+t}=0$. As a key source of
simplifications at $t=1$ this is not fully transferable to $t> 1$
\cite{Dubna} . At the higher $t$'s, one is simply expected to
pre-diagonalize the matrix ${\cal M}$ in textbook spirit. This is
a purely numerical step of course. In what follows we are going to
describe a more analytic approach. It will lie somewhere in
between the fully analytic (schematically, $t=0$) and purely
numerical (i.e., pre-diagonalization) extremes: Our unperturbed
propagators ${\cal R}$ will be constructed as sparse, triangular
matrices.

The presentation of this material will be split in three parts.
Firstly, using just $t=1$ for simplicity, subsection \ref{4.1}
explains the idea of constructing ${\cal R}$ in the upper
triangular form. Secondly, subsection \ref{4.2} employs the next,
$t=2$ example and explains the alternative approach using the
lower triangular unperturbed propagators.

We believe that this gives a clear guide to the general $t$'s.
Still, an abstract and detailed description of our innovated
perturbation theory (with any $t$) is offered in the Appendix. The
reason is that, building on the reader's experience with our
previous examples, we may introduce a less transparent but much
more compact notation. Moreover, we also relax there the immediate
connection of our technique to some peculiarities (e.g.,
one-dimensional nature) of our illustrative example (\ref{nase}).

\subsection{Upper triangular propagators:
$t=1$ example \label{4.1}}

We have split our eq.~(\ref{6cper}) into the separate rule
(\ref{fixing}), the upper part (\ref{7rfin}) with solution
(\ref{8fin}) and the lower part
\be
\left(
\begin{array}{ccccc}
a^{(0)}_{q+2}&d^{(0)}_{q+2} &&&\\
c^{(0)}_{q+3}&a^{(0)}_{q+3}&d^{(0)}_{q+3}  &&\\
&c^{(0)}_{q+4}&a^{(0)}_{q+4}&\ddots&
\\
&&\ddots&\ddots&d^{(0)}_{M-1}  \\
&&&c^{(0)}_{M}&a^{(0)}_{M}
\ea \right) \, \left( \ba
  h^{(k)}_{q+2}\\ h^{(k)}_{q+3}\\ h^{(k)}_{q+2}\\ \vdots\\ h^{(k)}_{M}
\ea \right) = \left( \ba
 \tilde{ \tau}^{(k-1)\star}_{q+2}\\ \tilde{
\tau}^{(k-1)}_{q+3}\\ \tilde{ \tau}^{(k-1)}_{q+4}\\ \vdots\\ \tilde{
\tau}^{(k-1)}_{M} \ea
\right) \label{6cperinf}
 \ee
truncated at certain $M \gg q+2$ and containing, on its right-hand
side, a re-defined, $^\star-$superscripted known quantity
$\tilde{\tau}^{(k-1)\star}_{ q+2} =\tilde{\tau}^{(k-1)}_{ q+2}
-c_{q+2}^{(0)}\,h^{(k)}_{q+1} $.

For a clear explanation of our main idea let us now drop the
superscripts and choose $q=1$ and $M=6$ again.  This returns us
back to our paedagogical example (\ref{llcdd}) re-written now in a
re-partitioned, equivalent square-matrix form
\be
\left(
\begin{array}{c|cccc}
c_3&a_3&d_3&&\\ &c_4&a_4&d_4&\\ &&c_5&a_5&d_5
 \\ &&&c_6&a_6
  \\ &&&&c_7
\ea \right) \, \left( \ba h_2\\ \hline h_3\\  h_4\\ h_5\\ h_6 \ea
\right) = \left( \ba
 \tilde{ \tau}_{3}\\
 \tilde{ \tau}_{4}\\
 \tilde{ \tau}_{5}\\
 \tilde{ \tau}_{6}\\
 c_7h_6
\ea
\right)
 \label{reproda}
 \ee
where $ h_2$ is already known and a trivial last row
$c_7h_6=c_7h_6$ has been added. The trick is that we may now
remove the cut-off completely. The infinite-dimensional
left-hand-side matrix
 \be {\cal Z}^{-1} = \left(
\begin{array}{ccccc}
c^{(0)}_{q+2}&a^{(0)}_{q+2}&d^{(0)}_{q+2} &&\\
&c^{(0)}_{q+3}&a^{(0)}_{q+3}&d^{(0)}_{q+3}  &\\
&&c^{(0)}_{q+4}&a^{(0)}_{q+4}&\ddots\\
 &&&\ddots&\ddots
\ea \right)  \label{prvni}
 \ee
is regular. It may be inverted in an algebraic, non-numerical and
cut-off-independent manner, ``forgetting" our use of the vectors
with $M=6$, i.e., $ \tilde{ \tau}_{7}=c_7h_6$ and $ h_{6+j}=0$ and
$\tilde{ \tau}_{7+j}=0$ for all $j = 1, 2, \ldots$. We may
conclude that once we know our upper triangular propagator matrix
${\cal Z}$ we may  pre-multiply by it equation (\ref{reproda})
from the left and get the final wave function defined by the
formula
\be
  h^{(k)}_{q+m}=
  \mu^{(k-1)}_{q+m}
+h^{(k)}_{M}\,c_{M+1}\,
  \nu^{(0)}_{q+m},
\ \ \ \ \ \ \ \ m = 1, 2, \ldots, M-q-1
 \label{wf}
 \ee
at any cut-off $ M \gg q+1$. Both its components have just an
elementary form
\be
\left( \ba
  \mu^{(k-1)}_{q+1}\\ \mu^{(k-1)}_{q+2}\\
\vdots\\ \mu^{(k-1)}_{M-2}\\ \mu^{(k-1)}_{M-1} \ea \right)= {\cal
Z}\, \left( \ba
 \tilde{ \tau}^{(k-1)}_{q+2}\\  \tilde{ \tau}^{(k-1)}_{q+3}\\ \vdots\\
 \tilde{ \tau}^{(k-1)}_{M-1}\\
 \tilde{ \tau}^{(k-1)}_{M}
\ea \right), \ \ \ \ \ \ \ \ \left( \ba
  \nu^{(0)}_{q+1}\\ \nu^{(0)}_{q+2}\\
\vdots\\ \nu^{(0)}_{M-2}\\ \nu^{(0)}_{M-1} \ea \right) =
\left( \ba
{\cal Z}_{1,M-q}\\
{\cal Z}_{2,M-q}\\
 \vdots\\
{\cal Z}_{M-q-2,M-q}\\
{\cal Z}_{M-q-1,M-q}
\ea \right).
\label{termini}
 \ee
This is a key point of our considerations. Starting from the first
omitted index $m=M-q$ our equation (\ref{wf}) is an identity. At
the first {\em admitted} (and exceptional) index $m=1$ this
equation defines, paradoxically, the right-hand side quantity
$h_M$ itself. Indeed, the pertaining left-hand side value $
h_{q+1}$ is already known.

In eq. (\ref{termini}) an {\em ascending} recurrent evaluations
may be recommended as giving, step by step, $\mu^{(k-1)}_{M-1} = {
\tilde{ \tau}^{(k-1)}_{M} / c_{M}^{(0)} } $ etc.  Our recipe is
complete.  We may summarize: In a deeply anharmonic double well
regime, the partial solvability of our unperturbed system {\em
and} the reducibility of its propagator to a triangular matrix
implies the feasibility of an innovated perturbation construction
with propagator ${\cal R}$ of an upper triangular matrix form.

An additional, marginal remark is due. If needed, the recurrences
(\ref{termini}) may be solved in a closed form
\be
\mu^{(k-1)}_{M-m-1}= {(-1)^m \over c^{(0)}_{M} c^{(0)}_{M-1}\ldots
c^{(0)}_{M-m}}\, F \ee where
\be
F= \det \left(
\begin{array}{cccccc}
 a^{(0)}_{M-m}&d^{(0)}_{M-m}&0&\ldots&0&
 \tilde{ \tau}^{(k-1)}_{M-m}
\\
c^{(0)}_{M-m+1}& a^{(0)}_{M-m+1}&d^{(0)}_{M-m+1}&&&
 \tilde{ \tau}^{(k-1)}_{M-m+1}
\\
0&\ddots&\ddots&\ddots&&\vdots\\ \vdots&\ddots&&&&\\ 0&&0&
c^{(0)}_{M-1}& a^{(0)}_{M-1}&
 \tilde{ \tau}^{(k-1)}_{M-1}\\
0&\ldots&&0& c^{(0)}_{M}&
 \tilde{ \tau}^{(k-1)}_{M}
\ea \right)  \label{perfidni}
 \ee
due to the Kramer's rule.

\subsection{Lower triangular propagators: $t=2$ example \label{4.2}}

Let us consider the pentadiagonal version of eq. (\ref{oureq}),
\ben \left(
\begin{array}{ccccccc}
a_0&d_0^{[1]} &d_0^{[2]} &0&0&0&\ldots\\
 c_1^{[1]}&a_1&d_1^{[1]}
&d_1^{[2]} &0&0&\ldots\\ c_2^{[2]}&c_2^{[1]}&a_2&d_2^{[1]} &0
&0&\ldots\\
 \hline  0&c_3^{[2]}&c_3^{[1]}&a_3&d_3^{[1]} &d_3^{[2]}
&\ldots\\
 \hline 0&0&c_4^{[2]}&c_4^{[1]}&a_4&d_4^{[1]}  &\ldots\\
\vdots&&&\ddots&\ddots&\ddots&\ddots \ea \right) \, \left( \ba
h_0^{(k)}\\  \left (h_1^{(k)}\right )\\  h_2^{(k)}\\
 h_3^{(k)}\\  h_4^{(k)}\\ \vdots
\ea \right)
 =
\left( \ba
  \tau_0^{(k-1)}\\ \tau_1^{(k-1)}\\ \tau_2^{(k-1)}\\ \hline
 \tau_3^{(k-1)}\\ \hline \tau_4^{(k-1)}\\ \vdots \ea \right)
 +E^{(k)} \
\left( \ba
  \rho_0^{\{0\}}
\\ \rho_1^{\{0\}}\\
 \rho_2^{\{0\}}\\ \hline \rho_3^{\{0\}}\\  \hline
  \rho_4^{\{0\}}\\
 \vdots \ea \right) \ \label{luycde}
 \een
with, say, $q=1$, i.e., $c_3^{[2]}=d_3^{[2]}=0$ and $
\rho_4^{\{0\}}=  \rho_5^{\{0\}}=\ldots=0$. After the insertion of
energy (\ref{knowne}) we sum the two right hand side vectors in
one, $\tau_j^{(k-1)}  +E^{(k)} \rho_j^{\{0\}} \equiv
\tilde{\tau}_j^{(k-1)}$, and omit the $_3-$subscripted row
(between lines). Simultaneously our normalization $h_q^{(k)}=0$
(in parenthesis) annihilates the second column in ${\cal M}$.

Moving further the first $t-1$ (i.e., one) plus one (exceptional,
$(q+t)-$th) columns of ${\cal M}$ to the right hand side and
dropping the redundant superscripts $^{(k)}$ and $^{(k-1)}$ we get
the equivalent equation
 \be
 {\cal Z}^{-1} \, \left( \ba
 h_2\\
 h_3\\  h_4\\  \hline
 h_6\\
 \vdots\\
  h_{M+2}\\
 \hline
 0\\
 \vdots
\ea \right)
 =
\left( \ba
  \tilde{\tau}_0\\
  \tilde{\tau}_1\\
   \tilde{\tau}_2\\
\hline
    \tilde{\tau}_4\\
\vdots \\
      \tilde{\tau}_M \\   \hline  \tilde{\tau}_{M+1}^\star
       \\ \vdots \ea \right)
 - h_0 \
\left( \ba
  \rho_0^{\{1\}}
\\ \rho_1^{\{1\}}\\
 \rho_2^{\{1\}}\\   \hline \rho_4^{\{1\}}\\
 \vdots \\
 \rho_M^{\{1\}} \\  \hline  \vdots \\ \vdots
 \ea \right)
- h_5 \
 \left( \ba
  \rho_0^{\{2\}}
\\ \rho_1^{\{2\}}\\
 \rho_2^{\{2\}}\\   \hline \rho_4^{\{2\}}\\
 \vdots \\
 \rho_M^{\{2\}} \\  \hline  \vdots\\ \vdots
 \ea \right)
  \ \label{lulcd}
 \ee
with the appropriate auxiliary $\tilde{\tau}_{M+1}^\star =
c_{M+1}^{[2]}\,h_{M-1}+ c_{M+1}^{[1]}\,h_{M}$ and
$\tilde{\tau}_{M+2}^\star = c_{M+2}^{[2]}\,h_{M}$ (while
$\tilde{\tau}_{M+3}^\star = \tilde{\tau}_{M+4}^\star = \ldots =
0$), with $ \rho_0^{\{1\}}=a_0$ etc, $\rho_4^{\{2\}}=d_4^{[1]}$
etc and with the lower triangular and infinite matrix
 \ben
 {\cal Z}^{-1}= \left(
\begin{array}{ccc|ccc}
d_0^{[2]} &&&&&\\ d_1^{[1]} &d_1^{[2]} &&&&\\
 a_2&d_2^{[1]} &d_2^{[2]} &&&\\
 \hline
 c_4^{[2]}&c_4^{[1]}&a_4&d_4^{[2]} &&\\
 0&c_5^{[2]}&c_5^{[1]}&d_5^{[1]} &d_5^{[2]}& \\
\vdots&\ddots&\ddots&\ddots&\ddots &\ddots\ea \right).
 \een
This pentadiagonal matrix is easily invertible since its main
diagonal is all non-zero. Indeed, by construction, $n \neq t+q$ in
$ d^{[t]}_n = B_t\,[\varepsilon_n-\varepsilon_{t+q} ]\,\langle
n|r^{2t}|n +t\rangle\ \label{elel}$. The action of ${\cal Z}$ upon
eq. (\ref{lulcd}) from the left gives our final wave functions
 \be
 \left( \ba
 h_2\\
 h_3\\  h_4\\  \hline
 h_6\\
 \vdots\\
  h_{M+2}
\ea \right)
 ={\cal Z}
\left( \ba
  \tilde{\tau}_0\\
  \tilde{\tau}_1\\
   \tilde{\tau}_2\\
\hline
    \tilde{\tau}_4\\
\vdots \\
      \tilde{\tau}_M
       \ea \right)
 - h_0 \ {\cal Z}
\left( \ba
  \rho_0^{\{1\}}
\\ \rho_1^{\{1\}}\\
 \rho_2^{\{1\}}\\   \hline \rho_4^{\{1\}}\\
 \vdots \\
 \rho_M^{\{1\}}
 \ea \right)
- h_5 \ {\cal Z}
 \left( \ba
  \rho_0^{\{2\}}
\\ \rho_1^{\{2\}}\\
 \rho_2^{\{2\}}\\   \hline \rho_4^{\{2\}}\\
 \vdots \\
 \rho_M^{\{2\}}
 \ea \right)
  \ \label{kilcd}.
 \ee
The seemingly redundant last two rows form in fact a core of the
whole construction: With the well known left-hand side values of
$h_{M+1}=h_{M+2}=0$ they must be read as the two necessary linear
algebraic equations needed to determine the two ``input
parameters" $h_0$ and $h_5$.

\subsubsection{An illustration}

A straightforward transition to the general $t$'s does not require
any new ideas but merely an appropriate ``shorthand" notation. Its
explicit description may be found in the Appendix. A nontrivial
advantage of its more abstract language is that one can
immediately work with a much more general class of Hamiltonians,
say, with their non-Hermiticity and asymmetry related to the
phenomenological absorption, etc. At the same time, even our
simple Pad\'e oscillators may provide a number of useful
applications.

Flexibility of our fairly weak assumptions may lead to a few
non-standard constructions. Imagine just an {\em interpolation}
between two zero-oder models.  Thus, our previous $t\geq 2$
solvable model (\ref{imple}) may be re-interpreted as the new
potential
\be
V_{[b]}(\delta,r) = r^2 + {[16+(1-\delta)\, \mu^{(1)}]\,r^2 + [2+
(1-\delta)\,\nu^{(1)}] \over 1-r^2 + r^4 } \label{implebex}
 \ee
also solvable at $\delta=0$. With a new ``small" parameter $\delta
= 1-\lambda$ and the new couplings
 \ben \mu^{(1)} = 14. 941 997\ldots - 16 \approx 1.05800,
\ \ \ \ \ \ \nu^{(1)} = 4. 959 146\ldots - 2 \approx 2.959 15
 \een
this leads to a methodically appealing {\em linear} interpolation
between the two solvable cases.

This may serve as a test of our method. In Table 6 the first-order
precision compares  well with the purely numerical exact energies.
In a way paralleling our above $t=1$ test (\ref{RSian}) the usual
evaluation of the $t=2$ overlap integrals is much more tedious of
course. Although the integrators in MAPLE \cite{Maple} still
manage and offer their evaluation, the immediate algebraic
solution of our linear algebraic three-by-three equations proves,
definitely, much preferable.

\section{Summary}

Our present recipe treats a phenomenological Schr\"{o}dinger
equation in three steps.  In the preliminary one we choose the
potential in its Pad\'{e} (or perturbed Pad\'{e}) asymptotically
harmonic representation (\ref{nase}). In the next preparatory
stage the rational potential with $2t+1$ free parameters is
assigned a suitable solvable zero-order approximant with as many
as $t+1$ free parameters.  We pick up the parity (or angular
momentum) $\ell$ and degree $q$ of the unperturbed wave function
and determine, algebraically, all the parameters which are
constrained by the solvability. In the third step we finally apply
our modified or innovated Rayleigh-Schr\"{o}dinger perturbation
algorithm. In each order we

\begin{itemize}

\item
compress our knowledge of preceding corrections (say, in an
auxiliary vector $\vec{\tau}$ given by formula (\ref{taurek}))
and, if needed,

\item
define immediately the new energy (by eq. (\ref{knowne}));

\item
construct another auxiliary array (in general, vector
$\vec{\theta}$ defined by recurrences (\ref{mourek}) in the
Appendix);

\item
choose a cut-off $M\gg 1$;

\item
satisfy the $M-$dependent model-space constraints (i.e., $t+1$
linear equations (\ref{four}) + (\ref{mour}) in general);

\item
evaluate all the missing components of the wave function
corrections (their general form is given by eq. (\ref{exoureq})
below) .

\end{itemize}

\noindent
As a comfortable methodical alternative to the current
prescriptions our innovated procedure admits a non-diagonality
of unperturbed Hamiltonians and avoids the necessity of their
pre-diagonalization at $\lambda=0$. Its computational efficiency
stems from its consequently recurrent character.

For practical puposes, it is promising that our recipe is
reducible, basically, just to a single recurrence relation per
each perturbation order.  This lowers the common danger of a
possible undetected numerical loss of precision, further
suppressed here (and especially in the context of non-linear
algebra in the zero-order constructions) by the high-precision
computer arithmetics and programming in MAPLE \cite{Maple}.

Originally, our choice of the illustrative rational potentials
(\ref{nase}) has been motivated, mathematically, by a comparative
smallness of their short-range perturbations. {\it A posteriori},
numerical tests clarified their phenomenological appeal. The
flexibility of their shapes proved paralleled by the ``fairly
dense" occurrence of their partially solvable bound states.
Indeed, their observables (e.g., energies) exhibit often an almost
linear or quadratic coupling-dependence. In a way extending the
$t=1$ observations by Gallas \cite{Gallas} this supports a very
good precision of perturbative predictions over a major part of
the coupling space. We may expect a facilitated tractability of
potentials with less common (e.g., multiple well) shapes.  The
frequent physical need of analysis of large variations of
realistic forces seems to have found here an adequate
computational tool.

\newpage

Table 1. Low-lying spectra in the four deepest solvable double
wells (\ref{fless}).

$$
\begin{array}{||c|c|cccc||}
\hline \hline \beta^{(0)}&q&\multicolumn{4}{c||}{\rm energies }\\
\hline 81.88      &                          3 & 18.999999996 &
19.  & 22.765764732 & 22.765764788 \\ && 26.526337990 &
26.526338492 & 30.281295324 & 30.281297900 \\ \hline
     64.89&                             3 & 17.  & 17.000000131 &
20.733677525 & 20.733679219 \\ && 24.459570379 & 24.459582299 &
28.176801248 & 28.176862466 \\ \hline 52.05& 3 & 15.301693677 &
15.301695784 & 18.999974785 & 19.  \\ & & 22.686594466 &
22.686760593 & 26.359595371 & 26.360391029 \\ \hline 49.91& 2 &
14.999996593 & 15.  & 18.690932465 & 18.690972685 \\ &&
22.369232872 & 22.369494489 & 26.032568100 & 26.033806209
\\
\hline 39.12& 3 & 13.356890687 & 13.356934267 & 17.  &
17.000474393 \\ & & 20.621974574 & 20.624839279 & 24.215073151 &
24.227701473 \\ \hline \hline \ea $$

\newpage

Table 2. Complete list of the $t=1$ roots $\beta^{(0)}$.

$$
\begin{array}{||c|c|c|c|c||}
\hline \hline q&{\rm parity}& \beta^{(0)}&\multicolumn{1}{c||}{\rm
excitation }&\multicolumn{1}{c||}{E^{(0)}}\\ \hline 0&{\rm even}&
6. &{\rm ground\ state}& 5.
\\ &{\rm odd}& 10.
                                   &{\rm ground\ state}&  7.
\\
\hline 1&{\rm even}& 8.8768943744  &{\rm first}&  9. \\ &&
17.123105626 &{\rm ground\ state}&  9.
\\&{\rm odd} & 12.   & {\rm first}& 11.
\\&
                                       & 26.  & {\rm ground\ state}& 11.
\\
\hline 2&{\rm even}& 11.490856174  & {\rm second}& 13.   \\& &
19.556337712 &{\rm first}&
 13.   \\& & 36.952806114 &{\rm ground\ state}&
13.
\\&{\rm odd} & 13.874580313 &{\rm second}& 15.
\\&
                                       & 28.206711029
&{\rm first}&  15.
\\&
                                       & 49.918708658 &{\rm ground\ state}&  15.
\\
\hline 3&{\rm even}& 13.816182739 &{\rm third}&  17.  \\& &
22.170398699 &
 {\rm second}& 17.   \\& & 39.118906994 &
{\rm first}&  17. \\& & 64.894511568 &{\rm ground\ state}&  17.
\\&{\rm odd} & 15.630566921& {\rm third}& 19.
\\&
                                       & 30.443898070 &{\rm second}&  19.
\\&
                                       & 52.049183356
& {\rm first}& 19.
\\&
                                       & 81.876351653 & {\rm ground\ state}& 19.
\\
\hline \hline \ea $$

\newpage

Table 3. $s-$wave roots $\mu^{(0)}$ and $\nu^{(0)}$ for potentials
(\ref{imple}) with $f=g=1$ and $q \leq 3$.

$$
\begin{array}{||c|cc|cccc||}
\hline \hline &\multicolumn{2}{c|}{\rm couplings}&
\multicolumn{4}{c||}{\rm coefficients}\\ q&
\mu^{(0)}&\nu^{(0)}&h_0^{(0)}&h_1^{(0)}&h_2^{(0)}&h_3^{(0)}\\
\hline 0&16&2&1&0&0&0\\ 1&14.9420&4.95915&-3.48195&1&0&0\\
2&14.0340&7.91914&8.18810&-3.79751&1&0\\
2&64.3015&-2.40610&1.02613&1.82256&1&0\\
3&13.2932&10.8801&-15.7092&9.19559&-3.91211&1\\
3&62.7170&0.883427&-1.93699&-2.48533&-0.0989786&1\\ \hline \hline
\ea $$

\newpage

Table 4.  $M \to \infty$ convergence of $s-$wave energies for
couplings (\ref{amp}) in (\ref{imple}).

$$
\begin{array}{||c|ccccccc||}
\hline \hline M&\multicolumn{7}{c||}{\rm low\ lying\  spectrum}\\
\hline 0 & 11.422198 &-&-&-&-&-&- \\ 1 & 10.672913 & 15.
&-&-&-&-&- \\ 2 & 10.945092 & 15.  & 16.2817 &-&-&-&- \\ 3 &
10.944852 & 15. &-&-&-&-&- \\ 4 & 10.944169 & 15.  & 18.5146
&-&-&-&- \\ 5 & 10.943697 & 15.  & 18.2954 & 20.8470 &-&-&- \\ 6 &
10.943435 & 15. & 18.1857 & 20.2068 & 24.0949 &-&- \\ 7 &
10.943317 & 15.  & 18.1222 & 20.0166 & 23.6752 & 28.065 &- \\ 8 &
10.943284 & 15.  & 18.0876 & 19.9294 & 23.5621 & 27.5340 & 34.6709
\\ 9 & 10.943292 & 15.  & 18.0717 & 19.8889 & 23.5096 & 27.4129 &
31.3782 \\ 10 & 10.943316 & 15. & 18.0670 & 19.8730 & 23.4838 &
27.3538 & 31.2387
\\ 11 & 10.943343 & 15.  & 18.0682 & 19.8704 & 23.4725 & 27.3221 &
31.1673 \\ 12 & 10.943366 & 15.  & 18.0718 & 19.8743 & 23.4696 &
27.3063 & 31.1255 \\ 13 & 10.943383 & 15.  & 18.0761 & 19.8806 &
23.4711 & 27.3002 & 31.1020 \\ 14 & 10.943395 & 15.  & 18.0798 &
19.8871 & 23.4745 & 27.2999 & 31.0907 \\ 15 & 10.943402 & 15.  &
18.0828 & 19.8928 & 23.4785 & 27.3027 & 31.0874
\\                                       16
                                       & 10.943406 & 15.  &
18.0849 & 19.8971 & 23.4822 & 27.3068 & 31.0889 \\ 17 & 10.943408
& 15.  & 18.0863 & 19.9002 & 23.4851 & 27.3110 & 31.0929 \\ 18 &
10.943408 & 15.  & 18.0872 & 19.9022 & 23.4874 & 27.3147 & 31.0977
\\ 19 & 10.943408 & 15.  & 18.0876 & 19.9034 & 23.4889 & 27.3177 &
31.1023 \\ 20 & 10.943408 & 15.  & 18.0878 & 19.9040 & 23.4899 &
27.3198 & 31.1083 \\ \hline \hline \ea $$

\newpage

Table 5. $t=3$ roots $u^{(0)}$, $v^{(0)}$ and $w^{(0)}$ for
 even parity,
 $q\leq 2$ and
  potentials (\ref{nubip}).

$$
\begin{array}{||r|rrr|rr||}
\hline \hline \multicolumn{1}{||c|}{\rm{\rm auxiliary\ root} } &
\multicolumn{3}{c|}{\rm couplings}& \multicolumn{2}{c||}{\rm
coefficients}\\ \multicolumn{1}{||c|}{  z} &w^{(0)} \ \ &v^{(0)} \
\ & u^{(0)} \ \ &h_0^{(0)} \  &h_1^{(0)}
\
\\
\hline \multicolumn{1}{||c|}{-} & 30.  & 0.  & 12.  & 1.  & 0.
\\
2.0534020780474 & 58.107 & -14.804 & 19.797 & 1.452 & 1.
\\
-11.95601933076 & 30.088 & 0.5698 & 15.691 & -8.454 & 1.
\\
3.9105205273467 & 93.642 & -33.229 & 33.945 & 1.091 & 2.258
\\
2.1316238057633 & 86.526 & 29.646 & 15.572 & 0.578 & 1.231
\\
-4.8208818863125 & 58.716 & -15.324 & 23.422 & -6.313 & -2.783
\\
-11.9379718900079 & 30.248 & 1.295 & 19.381 & 33.908 & -6.892
\\
\hline \hline \ea $$


\vspace{3truecm}

Table 6. Double test: Interpolation (\ref{implebex}) between the
two solvable potentials.

$$
\begin{array}{||c|c|c||}
\hline
  \hline {\rm ``large"\  coupling} &\lambda=1 & \delta=1-\lambda=1
\\ \hline
{\rm } &{\rm ground\ state} &{\rm first\ excitation}\\ \hline
\hline {\rm  unperturbed\ energy}& 11.0000&15.0000\\ {\rm the\
first\ correction}& -0.0449&-0.0985\\k=1 {\rm  \ approximation }&
10.9551&14.9015\\ \hline \hline {\rm Runge\ Kutta\ value}&
10.9434&14.6332\\ \hline \hline \ea $$

\newpage

\section*{Appendix. Perturbation construction
without left eigenvectors}

The main mathematical starting point of all our previous
considerations was the linear algebraic zero-order Schr\"{o}dinger
equation ${\cal M}(E^{(0)}) \vec{h}^{(0)}=0$ with elements such
that
 \ben
 {\cal M}_{m,m+t+j}=0, \ \ \ \ m = 0, 1, \ldots, \ \ j = 1,
2, \ldots
 \een
 \be
 {\cal M}_{n,n+t}=d^{[t]}_n>1/D >0, \ \ \ \ \ n = 0, 1, \ldots,
 \ \ \ \ n \neq n_0=t+q,
\label{domin}
 \ee
 \ben
 {\cal M}_{n_0,n_0+t} =d^{[t]}_{n_0}
  =1/D, \ \ \ \ \ \ 1/D \to 0.
 \een
Let us now forget about any one-dimensional interpretation or
Pad\'e-oscillator origin of these matrices which are, by
assumption, non-Hermitian. In such a case, it is natural to
suppose that the left eigenvector of ${\cal M}$ is
infinite-dimensional and ceases to be available. This appendix
will show that (and how) our formalism remains applicable even
without this auxiliary array.

\subsection*{A. Notation}

The mathematical importance of the upper-diagonal dominance
(\ref{domin}) in ${\cal M}$ lies in its relevance for recurrences.
Up to the exceptional $n=n_0$, each (i.e., the $n-$th) row of our
eq. (\ref{oureq}) may be read as an explicit recurrent definition
of its ``leftmost" unknown $h_{n}^{(k)}$. The exception is
unpleasant and its naive $^\ast-$superscripted regularization
\be
 d^{[t]\,\ast}_{n} =  \left\{
\begin{array}{ll}
D \neq 0\ \ & {\rm for}\ \ n =n_0= t+q\\ d^{[t]}_{n}\ \ & {\rm
otherwise} \ea \right. \label{naive}
 \ee
would produce a wrong value of $h_{2t+q}^{(k}$ without the
limiting transition $D \to \infty$. A slightly more sophisticated
recipe must be used. For its general formulation, the energy
$E^{(k)}$ and elements $\vec{h}^{(k)}$ will be split in a pair of
arrays $\vec{\zeta}$ and $\vec{\xi}$. The former vector will be
finite, containing just the $t+1$ ``difficult" components: energy
$ E^{(k)}\equiv\zeta_{0}$, initial values
${h}_j^{(k)}\equiv-\zeta_{j} $, $j = 1, 2,\ldots, t-1$ (let us now
prefer $h_0^{(k)}=0$ for normalization) and the exempted $
{h}^{(k)}_{2t+q}\equiv-\zeta_{t} $.  All the remaining unknowns
will lie in the infinite-dimensional vector $\vec{\xi}$ such that
$\xi_j = h_{j+t}^{(k)},\ j = 0, 1,\ldots$, $j \neq q$ (notice that
$j = q$ would double-count the exceptional $ {h}^{(k)}_{2t+q}$).

For the time being, let us leave the undetermined component of the
new vector $\vec{\xi}$ free. The presence of this new temporary
variable $z\equiv\xi_{t+q}$ opens the possibility of a
straightforward elimination of the vectors $\vec{\xi}$ as
functions of the {\em finite} number of unknowns in $\vec{\zeta}$.
Indeed, with an index $j$ out of the interval $j = 1, 2, \ldots,
t-1$ we may abbreviate all the $j-$th (i.e., leftmost) columns of
our zero-order matrix ${\cal M}$ as respective vectors
$\vec{\rho}^{\{j\}}$ distinguished by the braced superscript. We
move them all to the right-hand side of eq. (\ref{oureq}).  This
will reduce the left-hand side matrix ${\cal M}$ (acting on a
vector) to the mere triangular submatrix denoted as ${\cal Z}$
from now on (and acting on a subvector). After the above
simple-minded regularization (\ref{naive}) with
 \ben {\cal Z} \to {\cal Z}^\ast = \left(
\begin{array}{ccccc}
d_0^{[t]\,\ast }&0&0&0&\ldots\\ d_1^{[t-1] }&d_1^{[t]\,\ast
}&0&0&\ldots\\ d_2^{[t-2] }& d_2^{[t-1] }& d_2^{[t]\,\ast
}&0&\ldots\\ &&\ldots&& \ea \right)\
 \een
we may re-write our fundamental eq.  (\ref{oureq}) in the presence
of variable $z$,
\be
{\cal Z}^\ast\ \left( \ba
  \xi_{0}\\ \xi_{1}\\ \xi_{2}\\ \vdots
\ea \right) = \left( \ba
  \tau^{(k-1)}_{0}\\ \tau^{(k-1)}_{1}\\ \tau^{(k-1)}_{2}\\
\vdots \ea \right)+
 \sum_{j=0}^{t} \zeta_{j}
\left( \ba \rho^{\{j\}}_{0}\\ \rho^{\{j\}}_{1}\\
\rho^{\{j\}}_{2}\\ \vdots \ea \right)\ . \label{6cdpe}
\label{noureq}
 \ee
The last column $\vec{\rho} ^{\{t\}} $ must coincide with the
irregular, $2t+q-$th column of ${\cal M}$ without asterisk. The
new form of our equation carries more information than before. The
two auxiliary parameters (viz., $D$ and $z$) were introduced. It
easily follows from our construction that at any regularizing $D
\neq 0$, our new equation remains {\em precisely} equivalent to
its predecessor (\ref{oureq}) if and only if the redundant
variable $z$ vanishes, $z = 0$.  We may summarize: With the
necessary limiting transition $z \to 0$ deferred to the very end
of our considerations, the most important computational benefit of
the regularization $D\neq 0$ may be seen in a completely recurrent
solvability of our infinite-dimensional eq. (\ref{noureq}). Its
first row defines $\xi_0=h_t^{(k)}$ etc.

\subsection*{B. Reduction to a model space}

As long as, by construction, ${\rm det}\,{\cal Z}^\ast \neq 0$,
equation (\ref{noureq}) could be solved by an immediate matrix
inversion at an arbitrary $t$, $D$ and $z$, $\vec{\xi} = \left
({\cal Z}^\ast \right)^{-1 }\left(\vec{\tau}^{(k-1)} +
\sum_{j=0}^{t }\,\vec{\rho}^{\{j\}}\right)$.  This would specify
the infinite-dimensional left-hand side vector $\vec{\xi}$ as a
finite sum,
\be
\left( \ba
  \xi_{0}\\ \xi_{1}\\ \vdots
\ea \right) = \left( \ba
  \theta_{0}^{(k-1)}\\ \theta_{1}^{(k-1)}\\ \vdots
\ea \right)+
 \sum_{j=0}^{t}\ \zeta_{j}\
\left( \ba
  \eta^{\{j\}}_{0}\\ \eta^{\{j\}}_{1}\\
\vdots \ea \right)\ . \label{exoureq} \ee Accepting this idea, we
may even try to determine each of the right-hand side components
separately.  This would be a tremendous simplification of the
algorithm since the last $t+1$ individual recurrences are
order-independent,
\be
\left(
\begin{array}{cccc}
d_0^{[t]\,\ast }&0&0&\ldots\\ d_1^{[t-1]
}&d_1^{[t]\,\ast}&0&\ldots\\ &&\ldots& \ea \right) \ \left( \ba
  \eta_{0}^{\{j\}}\\ \eta_{1}^{\{j\}}\\ \vdots
\ea \right) = \left( \ba
  \rho^{\{j\}}_{0}\\ \rho^{\{j\}}_{1}\\
\vdots \ea \right), \ \ j = 0, 1, \ldots, t. \label{fourek} \ee
They just ``reparametrize"  a part of our zero-order Hamiltonian.
Only the very first definition will vary with the order $k$,
\be
\left(
\begin{array}{cccc}
d_0^{[t]\,\ast }&0&0&\ldots\\ d_1^{[t-1]
}&d_1^{[t]\,\ast}&0&\ldots\\ &&\ldots& \ea \right) \  \left( \ba
\theta_{0}^{(k-1)}\\ \theta_{1}^{(k-1)}\\ \vdots \ea \right) =
\left( \ba
  \tau_{0}^{(k-1)}\\ \tau_{1}^{(k-1)}\\ \vdots
\ea \right)\ . \label{mourek} \ee Still, it depends neither on the
unknown energy $E^{(k)}$ nor on the unknown coefficients
$\vec{h}^{(k)}$. We may say that it ``compactifies" the previous,
known results $E^{(k-1)}$, $\vec{h}^{(k-1)}$, $ E^{(k-2)},\
\ldots$, and ``compresses" them into a new input vector
$\vec{\theta}^{ (k-1)} =\left({\cal Z}^\ast\right)^{-1}\, \vec
{\tau}^{(k-1)} $.

It remains for us to find the values of the $t+1$ unknown
parameters $\zeta_j$.  For this purpose we return to the
``forgotten" truncation conditions
\be
\xi_{M+1} =\xi_{M+2} =\ldots =\xi_{M+t} = 0 \label{experi}.
 \ee
In the present notation these requirements read
\be
\ba \theta_{M+1}^{(k-1)}  + \sum_{j=0}^{t} {\eta}^{\{j\}}_{M+1}
\zeta_{j} = 0\\ \theta_{M+2}^{(k-1)}  + \sum_{j=0}^{t}
{\eta}^{\{j\}}_{M+2} \zeta_{j} = 0\\ \ldots\\ \theta_{M+t}
^{(k-1)} + \sum_{j=0}^{t} {\eta}^{\{j\}}_{M+t} \zeta_{j} = 0\ \ea
\label{four} \ee and suppress the variability of
parameters~$\vec{\zeta}$.  By construction, all these equations
still depend on our auxiliary and, generically, nonvanishing
variable $z$.  Vice versa, the validity condition $z=0$ may only
be re-introduced as an additional, explicit requirement $\xi_{t+q}
= 0$, i.e., as the $(t+1)-$st equation
\be
\theta_{t+q}^{(k-1)}  + \sum_{j=0}^{t} {\eta}^{\{j\}}_{t+q}
\zeta_{j} = 0\ . \label{mour} \ee The concatenated system
(\ref{mour}) + (\ref{four}) of $t+1$ conditions is our final
model-space-like formula.  It defines all the $t+1$ parameters
collected in the array $\vec{\zeta}$.

At the correct value of energy (\ref{knowne}) our auxiliary
regularization variable $z$ becomes identically equal to zero and
the redundant equation (\ref{mour}) may be omitted. Our
model-space-like boundary conditions then degenerate to the mere
$t$ equations~(\ref{four}).

\subsection*{C. Illustrations}

In the real fixed-point arithmetics an accumulation of errors may
make the numerical value of $z$ (i.e., the right-hand side of eq.
(\ref{mour})) still slightly different from zero. Nevertheless, as
long as the components of the latter equation itself are all of
the order ${\cal O}(D^{-1})$ for large $D \gg 1$ (cf. eq.
(\ref{inver})) the errors accumulate in the product $Dz$ rather
than in the quantity $z$ itself.  The choice of a sufficiently
large $D\to \infty$ settles the problem of errors.

\subsubsection*{C.1. $t=1$}

After we return to $t=1$ for illustration, the last line of eq.
(\ref{kkdd}) will represent our $z=0$ constraint (\ref{mour}).
Similarly, the last line of eq. (\ref{llcdd}) plays the role of
the second constraint (\ref{four}).  Once we define the auxiliary
two-column matrix $\eta$, \ben \vec{\eta}^{\{0\}} = ({\cal Z}^\ast
)^{-1} \left( \ba
     \rho^{\{0\}}_0\\ \rho^{\{0\}}_1\\
\vdots\\ \rho^{\{0\}}_{q+1} \ea \right), \ \ \ \ \ \ \ \
\vec{\eta}^{\{1\}} = ({\cal Z}^\ast )^{-1} \left( \ba 0\\ \vdots\\
0\\ a_{q+2}(0)\\ c_{q+3}(0) \ea \right)\
 \een
the inverse matrix in the pair of equations (\ref{mour}) and
(\ref{four}) remains triangular,
\be
\left( \ba -E^{(k)}\\ h_{q+2}^{(k)} \ea \right)
=
\left(
\begin{array}{cc}
{\eta}^{\{0\}}_{q+1}& 0\\ {\eta}^{\{0\}}_{M+1}&
{\eta}^{\{1\}}_{M+1} \ea \right)^{-1} \left( \ba
\theta^{(k-1)}_{q+1}\\ \theta^{(k-1)}_{M+1} \ea \right), \ \ \ \ \
\ M \gg q \geq 0\ , \ \ \ \ t=1. \label{zustane}
 \ee
This means that in any order $k$ and at an arbitrary termination
$q$ and/or cut-off $M<\infty$ the definition of energies $E^{(k)}$
remains finite-dimensional.

\subsubsection*{C.2. $t=2$}

Above, we have already chosen $q=1$ and $M=6$ for illustration at
$t=2$. Here, this gives the triangular regularized submatrix
${\cal Z}^\ast$ which has the ordinary square-matrix form
 \ben {\cal Z}^\ast = \left(
\begin{array}{ccc|c|ccc}
d_0^{[2]}&0&0&0&0&0&0
\\ d^{[1]}_1&d_1^{[2]}&0 &0&0&0&0\\
a_{2}&d^{[1]}_{2}&d^{[2]}_{2}&0&0&0&0\\ \hline
c^{[1]}_{3}&a_{3}&d^{[1]}_{3}&D&0&0&0\\ \hline
c^{[2]}_{4}&c^{[1]}_{4}&a_{4}&d^{[1]}_{4}&d^{[2]}_{4}&0&0\\
0&c^{[2]}_{5}&c^{[1]}_{5}&a_{5}&d^{[1]}_{5}&d^{[2]\,\star}_{5}&0\\
0&0&c^{[2]}_{6}&c^{[1]}_{6}&a_{6}&d^{[1]\,\star}_{6}&d^{[2]\,\star}_{6}
\ea \right) \ , \ \ \ \ t=2.
 \een
Its elements marked by the superscript$^\star$ are in fact
arbitrary but, wherever possible, we shall keep them strictly
equal to their nonzero unmarked values.  Such a convention (with
$d^{[2]\,\star}_{5}= d^{[2]}_{5}$, $d^{[2]\,\star}_{6}=
d^{[2]}_{6}$ and $d^{[1]\,\star}_{6}= d^{[1]}_{6}$) makes our
construction less cut-off-dependent. Only the single element
$d^{[2]\,\star}_{3}\equiv D\neq 0$ remains really indeterminate.

\subsubsection*{C.3. $D \approx 0$}

Whenever the value of $D$ vanishes in a ``correct" limit $D \to
0$, matrix ${\cal Z}$ drops its asterisk and equation
(\ref{oureq}) acquires its non-recurrent, pseudo-inversion form,
\be
{\cal Z}\ \left( \ba h_2^{(k)}\\ h_3^{(k)}\\ h_4^{(k)}\\ \hline
0\\ \hline
 h_6^{(k)}
\\ h_7^{(k)\,\star}
\\ h_8^{(k)\,\star}
\ea \right)
 =
\left( \ba
  \tau_0^{(k-1)}\\ \tau_1^{(k-1)}\\ \tau_2^{(k-1)}\\
\hline
 \tau_3^{(k-1)}\\
\hline \tau_4^{(k-1)}\\ \tau_5^{(k-1)}\\ \tau_6^{(k-1)} \ea
\right)
 +E^{(k)} \
\left( \ba
  \rho_0^{\{0\}}\\ \rho_1^{\{0\}}\\ \rho_2^{\{0\}}\\
\hline \rho_3^{\{0\}}\\ \hline 0 \\ 0 \\ 0 \ea \right) -
h_1^{(k)}\ \left( \ba d_0^{[1]}\\ a_1\\ c^{[1]}_2\\ \hline 0\\
\hline
 0\\ 0\\0
\ea \right) - h_5^{(k)} \ \left( \ba 0\\ 0\\ 0\\ \hline 0\\ \hline
d_4^{[1]}\\ a_5\\ c^{[1]}_6 \ea \right)\ . \label{lepen} \ee Step
by step, it defines the ``upper" coefficients $h_2^{(k)}$,
$h_3^{(k)}$ and $h_4^{(k)}$ as functions of the energy {\em and}
of another undetermined parameter $h_1^{(k)}$. Similarly, the
first ``lower" nonzero coefficient $h_6^{(k)}$ is specified as a
quantity which depends on all the ``upper" coefficients {\em plus}
on a new parameter $h_5^{(k)}$.  At the end, the correct values of
our three unknown variables should be fixed by the ``forgotten"
fourth row and by the two truncation conditions
$h_7^{(k)^\star}=h_8^{(k)^\star}=0$.

\subsubsection*{C.4. $D \gg 0$}

In the example (\ref{lepen}), the regularization ${\cal Z }(D=0)
\to {\cal Z}^\ast(D \neq 0)$ leads to the equivalent eq.
(\ref{noureq}) if and only if the contributions proportional to
$z$ vanish, $z \to 0$.  The situation remains virtually unchanged
when we admit a growth of $q$.  We may invert, in partitioned
notation,
\be
\left( {\cal Z}^\ast \right)^{-1} = \left(
\begin{array}{c|c|c}
{\cal Z}_1&0&0\\ \hline u^T&D&0\\ \hline {\cal Z}_3 &w& {\cal Z}_2
\ea \right)^{-1}= \left(
\begin{array}{c|c|c}
{\cal Z}_1^{-1}&0&0\\ \hline -D^{-1}u^T {\cal Z}_1^{-1}
&D^{-1}&0\\ \hline {\cal Z}_2^{-1} {\cal C} {\cal Z}_1^{-1}
&-{\cal Z}_2^{-1}wD^{-1} & {\cal Z}_2^{-1} \ea \right)
\label{inver}
 \ee
(with $ {\cal C} = -{\cal Z}_3 +D^{-1}wu^T $) and define the three
order-independent and infinite-dimensional vectors
\be
\left( \vec{\eta}^{\{0\}},\ \vec{\eta}^{\{1\}},\
\vec{\eta}^{\{2\}}\right)
=
\left( {\cal Z}^\ast \right)^{-1}\left[ \left( \ba
\rho^{\{0\}}_0\\ \vdots\\ \rho^{\{0\}}_{q+1}\\
\rho^{\{0\}}_{q+2}\\ 0\\ \vdots\\ 0 \ea \right), \left( \ba
 d_0^{[1]}\\
\vdots\\ c^{[2]}_3\\0\\ \vdots\\ \vdots\\ 0 \ea \right), \left(
\ba 0\\ \vdots\\ \vdots\\ 0\\ d_{q+3}^{[1]}\\ \vdots\\
c^{[2]}_{q+6} \ea \right) \right]\ . \label{diesirae}
 \ee It is an
easy linear algebra to show that the $D-$dependence of these
components of the wavefunction corrections (cf. eqs.
(\ref{noureq}) and (\ref{inver})) implies their linear
$z-$dependence. We may choose $D \gg 1$ and get the small
numerical span of $z = {\cal O}(D^{-1})$, i.e., only a small
spuriosity in our tentative $z \neq 0$ wavefunctions, $
h_j^{(k)}(z)= h_j^{(k)}(0)+{\cal O} (z) $.  In the limit $D \to
\infty$ we just return to the formulae studied above.
Nevertheless, also any time {\em before} such a limiting
transition, equations (\ref{mour}) and (\ref{four}) define our
last unknown parameters $\vec{\zeta}$ in terms of the vectors
(\ref{diesirae}), via the three-dimensional matrix inversion
\be
 \left(
\ba -E^{(k)}\\ h_{1}^{(k)}\\ h_{q+4}^{(k)} \ea \right)
=
\left(
\begin{array}{ccc}
{\eta}^{\{0\}}_{q+2}& {\eta}^{\{1\}}_{q+2}& 0\\
{\eta}^{\{0\}}_{M+1}& {\eta}^{\{1\}}_{M+1}& {\eta}^{\{2\}}_{M+1}\\
{\eta}^{\{0\}}_{M+2}& {\eta}^{\{1\}}_{M+2}& {\eta}^{\{2\}}_{M+2}
\ea \right)^{-1}\, \left( \ba \theta^{(k-1)}_{q+2}\\
\theta^{(k-1)}_{M+1}\\ \theta^{(k-1)}_{M+2} \ea \right), \ \ \ \ \
\ M \gg q \geq 0. \ee The occurrence of a zero matrix element is a
peculiarity of the scheme (cf. eq. (\ref{diesirae})).

\newpage

\section*{Figure captions}

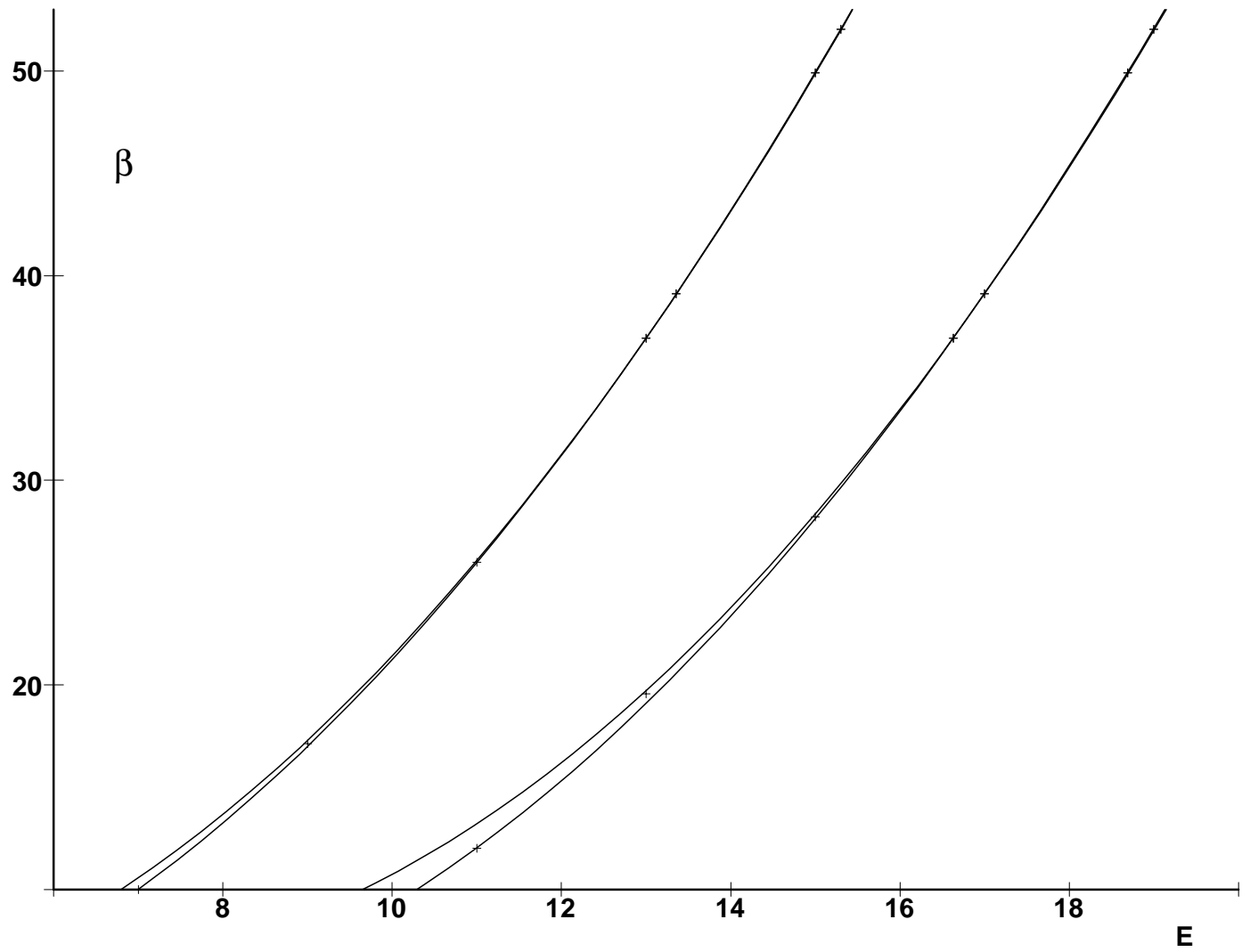
\noindent Figure 1. Coupling $\beta$ vs. energy $E$ for the first
four bound states in potential (\ref{fless}) with $B=1$.

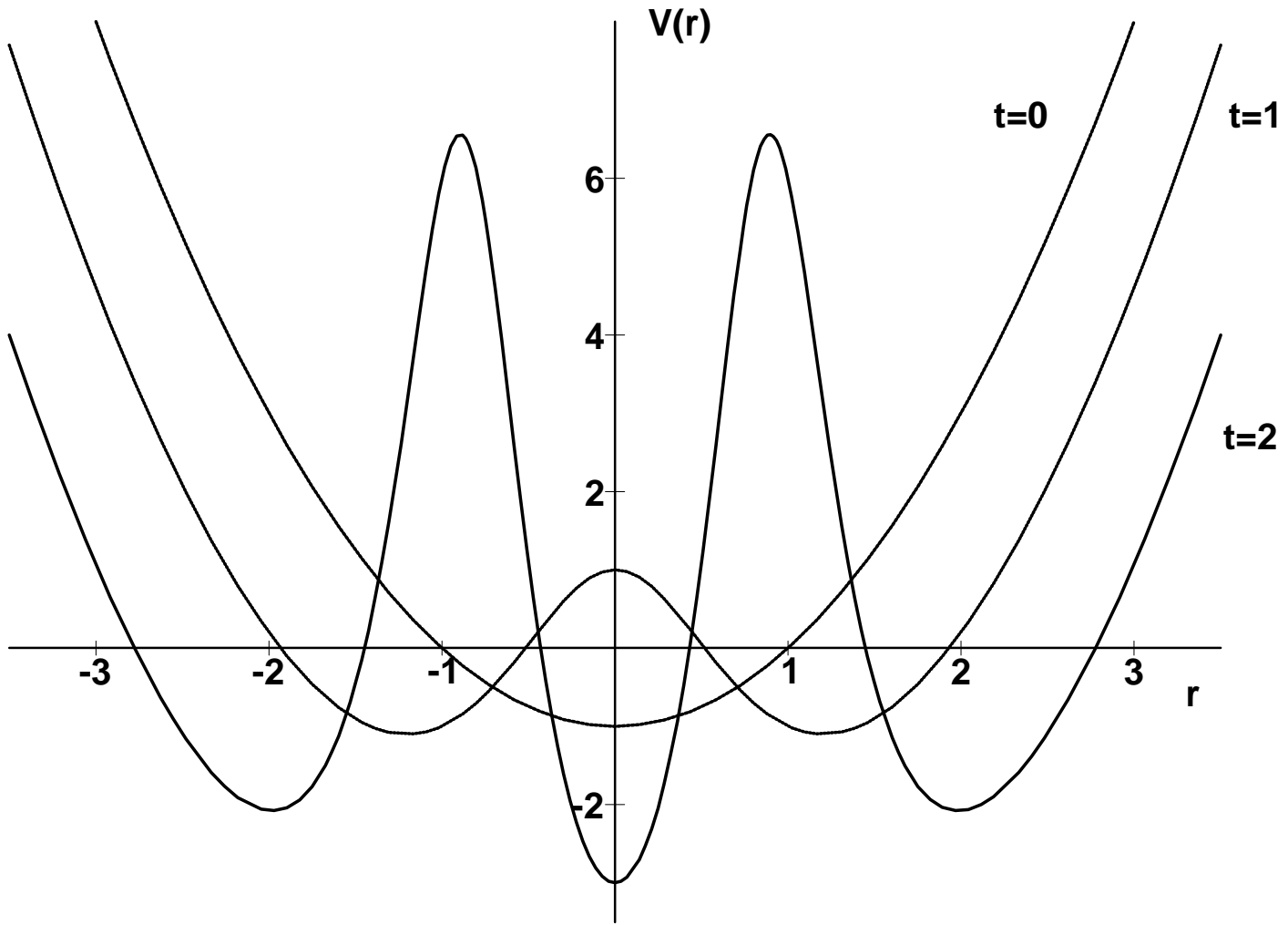
\noindent Figure 2. Three potentials (\ref{nase}) supporting the
exact ground state at zero energy.

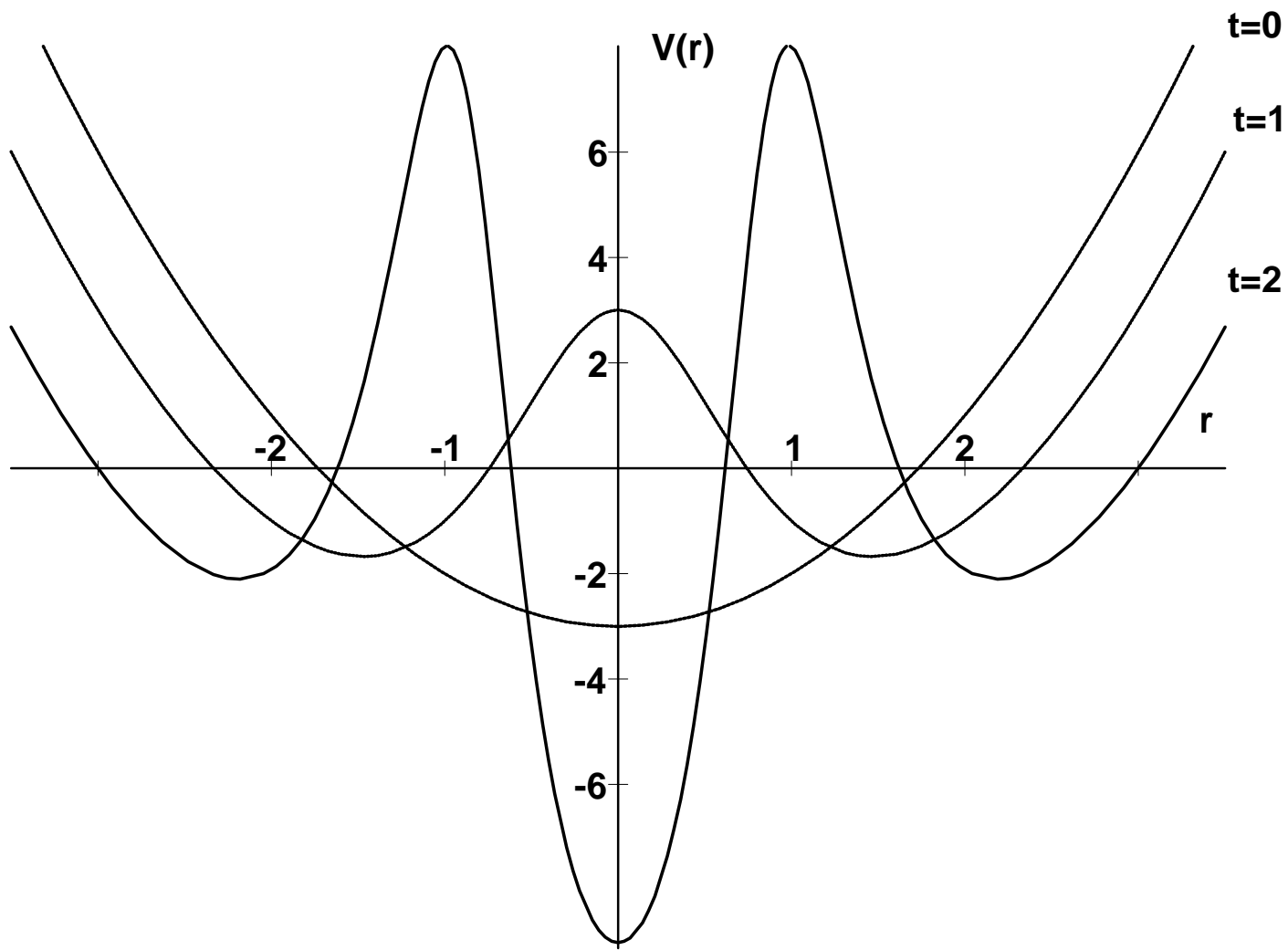
\noindent Figure 3. Three potentials (\ref{nase}) supporting the
exact first excited state at zero energy.

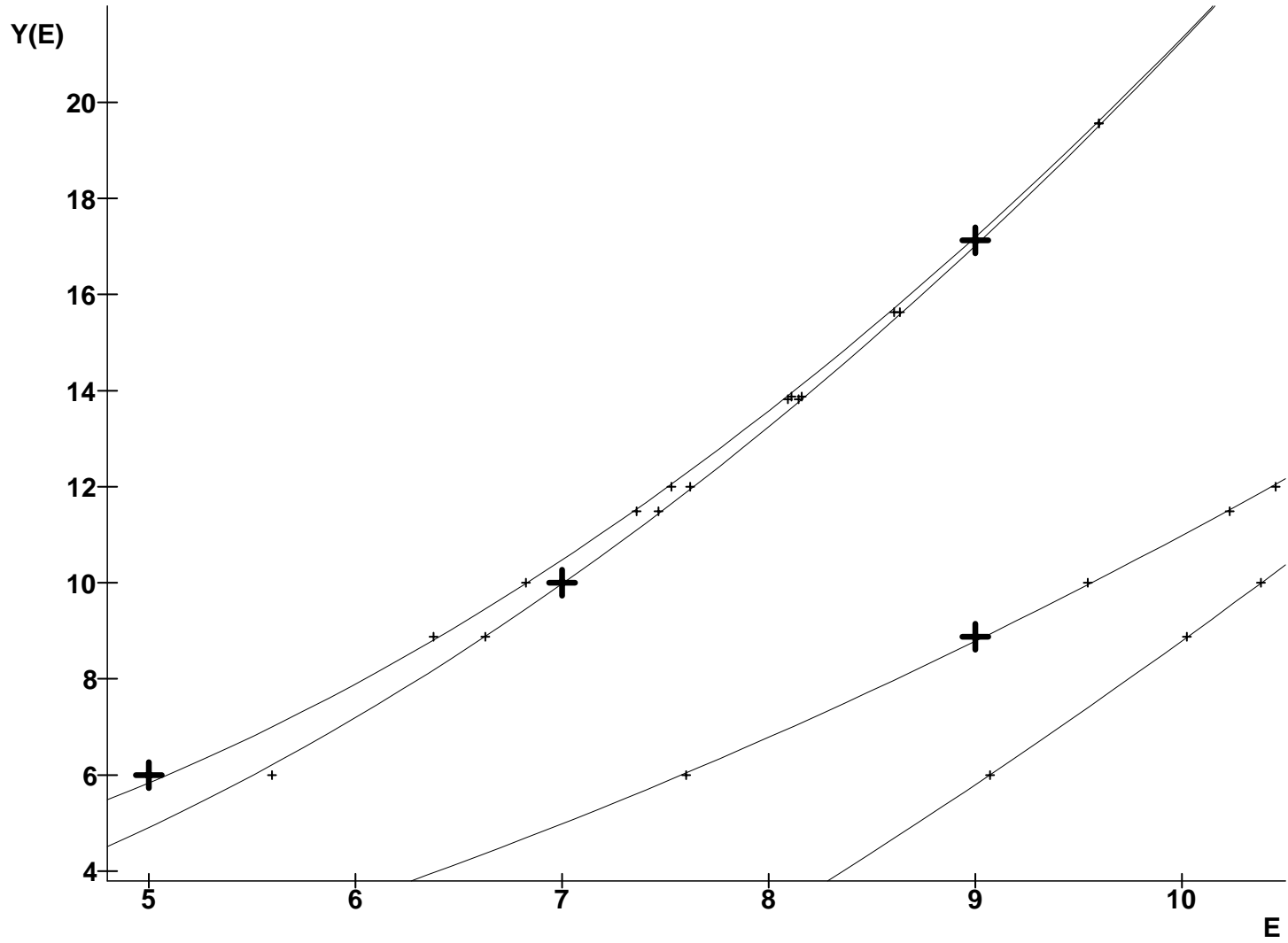
\noindent Figure 4. An empirical fit $Y(E)=a+bE+cE^2$ of the
couplings $\beta=\beta(E)$. The thin and thick crosses denote the
respective numerical and non-numerical levels at various
$\beta^{(0)}$.

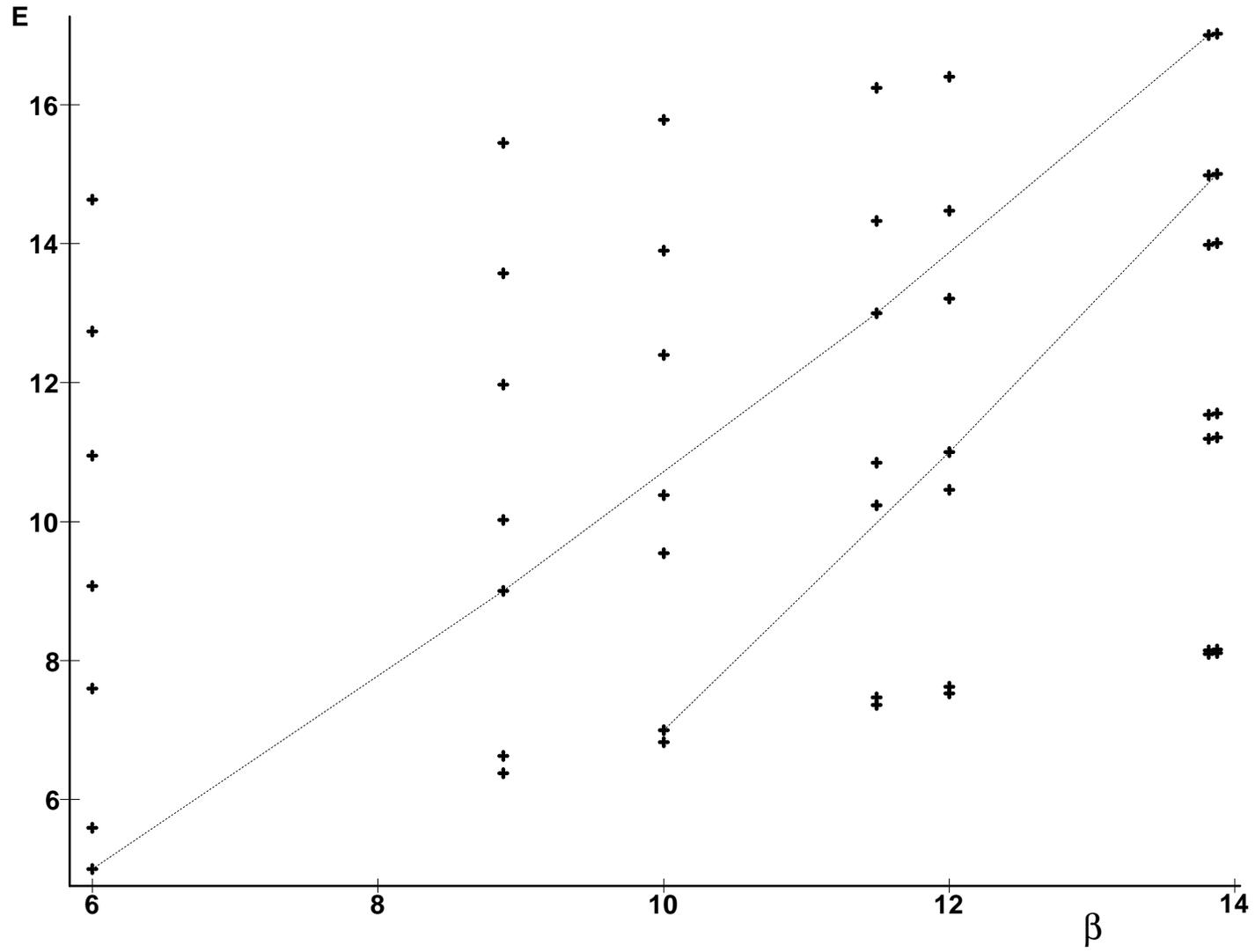
\noindent Figure 5.  The seven lowest energies for the seven
lowest barriers $\beta^{(0)}$ in (\ref{fless}).  The auxiliary
``Gallas" lines connect the symmetric (upper curve) and asymmetric
(lower curve) exact levels with growing $q$.

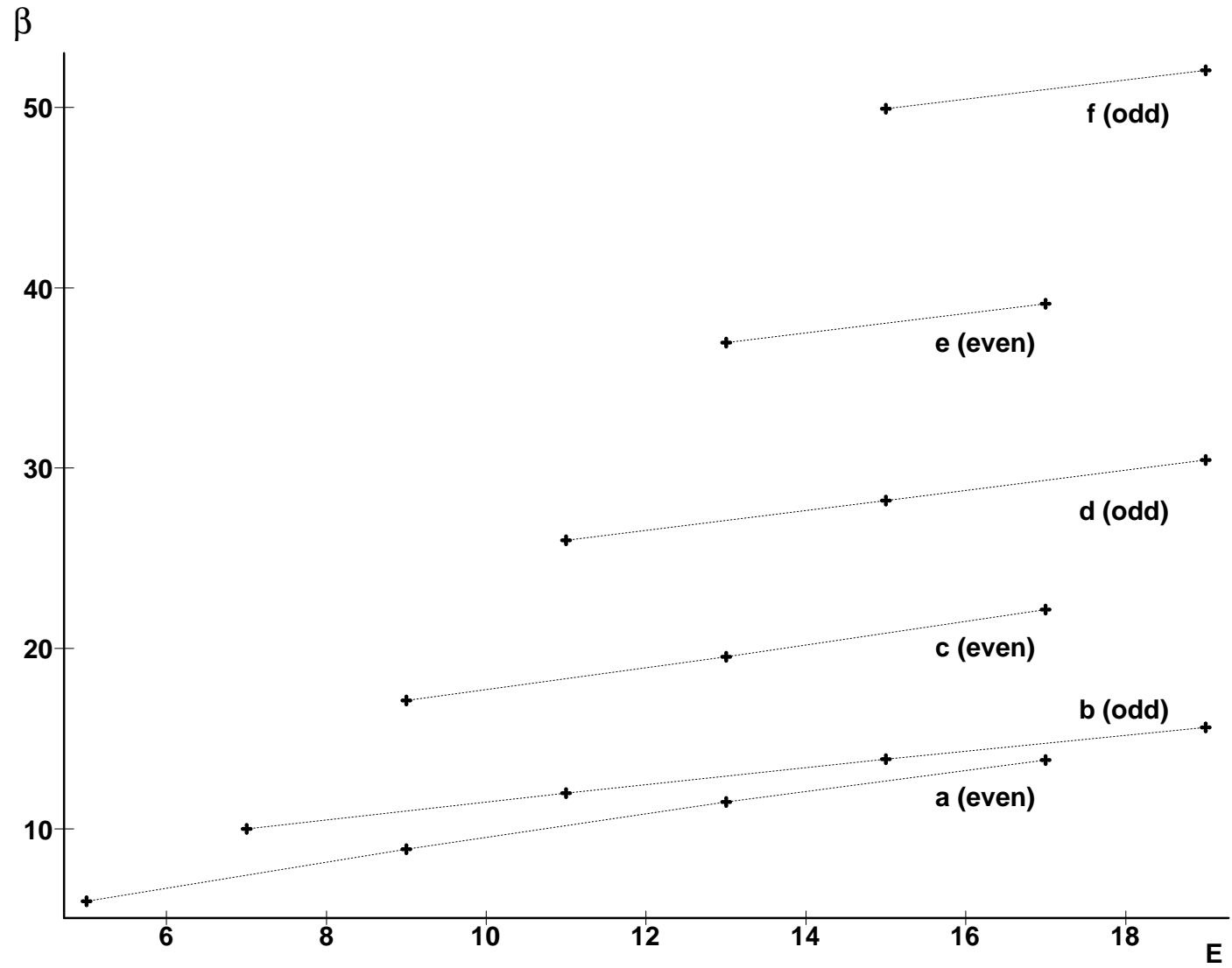
\noindent Figure 6. Several ``Gallas" lines in a bigger part of
the $E - \beta$ plane.


\newpage

\begin{thebibliography}{00}

\bibitem{BWu}
T. Kunihiro, Phys. Rev. D 57 (1998), R2035, with further
references.

\bibitem{Dyson}
F. J. Dyson, Phys. Rev. 85 (1952), 631.

\bibitem{Simon}
B. Simon, Int. J. Quant. Chem. 21 (1982), 3, with a concise
historical review.

\bibitem{Turbiner}
A. V. Turbiner and A. G. Ushveridze, J. Math. Phys. 29 (1988),
2053.

\bibitem{shortrev}
A. K. Mitra, J. Math. Phys. 19 (1978), 2018;

S. R. Kaushal, J. Phys. A: Math. Gen. 12 (1979), L253;

J. Killingbeck and S. Galicia, Phys. Lett. A 71 (1979), 17;

A. Hautot, J. Comput. Phys. 39 (1981), 72;

C. S. Lai and H. E. Lin, J. Phys. A: Math. Gen. 15 (1982), 1495;

Y. P. Varshni, Phys. Rev. A 36 (1987), 3009.

\bibitem{flessi}
H. Risken and H. Vollmer, Z. Phys. 201 (1967), 323;

G. P. Flessas, Phys. Lett. A 83 (1981), 121;

V. S. Varma, J. Phys. A: Math. Gen. 14 (1981), L489;

J. Heading, J. Phys. A: Math. Gen. 15 (1982), 2355;

M. H. Blecher and P. G. L. Leach, J. Phys. A: Math. Gen. 20
(1987), 5923;

M. Znojil and P. G. L. Leach, J. Math. Phys. 33 (1992), 2785;

A. G. Ushveridze, Quasi-exactly Solvable Models in Quantum
Mechanics, IOP Publishing, 1994.

\bibitem{jasi}
M. Znojil, Phys. Rev. A 35 (1987), 2448;

M. Znojil, J. Math. Phys. 29 (1991), 2611;

M. Znojil, Phys. Lett. A 155 (1991), 87.

\bibitem{Dubna}
M. Znojil, A new perturbative treatment of pentadiagonal
Hamiltonians, Report E4-87-655, JINR Dubna, 1987;

M. Znojil, Czechosl. J.  Phys. 41 (1991), 397 and 497.

\bibitem{Bes}
N. Bessis and G. Bessis, J. Math. Phys. 21 (1980), 2780.

\bibitem{jasib}
R. F. Bishop, M. F. Flynn  and M. Znojil, Phys. Rev. A 39 (1989),
5336.

\bibitem{Mitra}
R. N. Chaudhuri and B. Mukherjee, J. Phys. A: Math. Gen. 16
(1983), 4031;

M. Cohen, J. Phys. A: Math. Gen. 17 (1984), 2345;

C. Handy, J. Phys. A: Math. Gen. 18 (1985), 3593;

P. Roy and R. Roychoudhury, Phys. Lett. A 122 (1987), 275;

H. Scherrer, H. Risken and J. Leiber, Phys. Rev. A 38 (1987),
3949.

\bibitem{Whitehead}
R. R. Whitehead, A. Watt, G. P. Flessas and M. A. Nagarajan, J.
Phys. A: Math. Gen. 15 (1982), 1217.

\bibitem{comm}
M. Znojil, J. Math. Chem. 19 (1996), 205.

\bibitem{cf}
M. Znojil, J. Phys. A: Math. Gen. 16 (1983), 293;

F. M. Fern\'{a}ndez, Phys. Lett. A 160 (1991), 116;

R. K. Agrawal and V. S. Varma, Phys. Rev. A 48 (1993), 1921;

C. Stubbins and M. Gornstein, Phys. Lett. A 202 (1995), 34.

\bibitem{Sturmians}
M. Znojil, J. Math. Phys. 38 (1997) 5087.

\bibitem{Korn}
A. G. Korn and T. M. Korn, Mathematical Handbook, McGraw-Hill,
1968, ch. 1.7-4.

\bibitem{Gallas}
J. A. C. Gallas, J. Phys. A: Math. Gen. 22 (1988,) 3393;

A. Lakhtakia, J. Phys. A: Math. Gen. 23 (1989), 1701;

G. Vanden Berghe and H. De Meyer, J. Phys. A: Math. Gen. 23
(1989), 1705;

M. Znojil, in Selected Topics in QFT and Mathematical Physics,
eds. J. Niederle and J. Fischer, World Scientific, 1990, p. 376.

\bibitem{Kato}
T. Kato, Perturbation Theory for Linear Operators, Springer, 1966.

\bibitem{Maple}
B. W. Char et al, First leaves: A Tutorial Introduction to Maple
V, Springer, 1992.

\end{thebibliography}
\end{document}